# Possibilities, challenges and limits of a European charters corpus (CEMA)

Nicolas Perreaux, LaMOP[1]

"*He who has a thousand francs buys himself a bicycle: he has enough for life. He who has a million, when he has bought himself a bicycle every year for twenty years, is on top of the world. But he who can buy ten thousand bicycles a second for ten thousand years – he is afraid.*"[2] What is the link between this phobia of large numbers and the study of ancient texts? In 2021, the state of digitised medieval documentary corpora, and even more so of diplomatic corpora, might seem paradoxical. The latter have in fact existed in large numbers for several decades now. Medievalists in particular have been great 'digitizers' since the beginning of the second half of the 20th century. It is in this discipline that the largest digital corpora of ancient documents are available, on the one hand because the corpus from Antiquity is much smaller, and on the other hand because corpora from the 16th-18th centuries are little edited and therefore little digitised. Nevertheless, these medieval textual collections remain generally untapped.

Let us take a few examples: it is well known that the first corpus of digitised ancient texts is the *Corpus Thomisticum*, whose digitisation by Roberto Busa began in 1949[3]. That's more than 70 years ago! To date, however, the database has not been the subject of any quantitative, stylistic or even general semantic study. This situation is in fact almost universal: attempts to systematically exploit large textual corpora in history are rare. To my knowledge, the *Corpus Inscriptionum Latinarum*, the first volume of which was published in 1893 and which is now largely digitised[4], has never been the subject of an attempt at global manipulation. The same is true of the *Corpus Christianorum*, the AASS (*Acta Sanctorum*) and many others. How can we explain this contradiction between our capacity to accumulate corpora and analyse them? Should we deduce that these "tools" are ultimately of limited interest? On the contrary, it seems that it is precisely because these databases constitute a paradigmatic revolution – in the sense that Thomas Kuhn understood this term in 1962[5] – that they have been and still are underused. This relative sluggishness is not due to the training of historians (who could very well be trained in digital tools), or even to technological developments. The historical method as it has been conceived since the 17th-18th centuries is in fact particularly effective when it comes to analysing a limited group of occurrences[6].

However, databases pose the documentary question in a significantly different way, because they offer the possibility of exploiting complete corpora, varying the sets of scales, but also exploiting much larger sets[7]. What can we do, for example, with the 500,000 mentions of *terra* that we find in medieval Latin databases, or with the 700,000 instances of *sanctus* (excluding *Acta Sanctorum*)? To explore one of the possibilities, the objective of this paper is to present a meta-corpus of diplomatic documents entitled *Cartae Europae Medii Aevi* or CEMA. I will try to show the logic and limits of this meta-corpus, by specifying both its structure and its possible future extensions. The second part of the paper will be devoted to specific examples that will attempt to show the interest of such a database. The third part will examine the possibilities opened up by the corpus in terms of historical semantics.

[1] **This text is a draft**, whose objective is to provide a translation to the video presentations of the CEMA, given on different occasions (in particular during the seminar H-37, directed by Paul Bertrand, at the University of Louvain; and during the seminar « Sciences du patrimoine - sciences du texte. Confrontation des méthodes », directed by Thibault Clérice and Elsa Marguin-Hamont, at the École nationale des Chartes). For this reason, I have kept footnotes to a minimum. Full bibliographies are available in the articles referenced in the case presentations. Also, I would like to thank Anna Dorofeeva for her kind review of the English version of the draft.

[2] *Crésus*. Dir. Jean Giono, Les Films Jean Giono, 1960. In French: « Celui qui a mille francs, il s'achète une bicyclette : il en a pour la vie. Celui qui a un million, quand il se sera acheté une bicyclette toutes les années pendant vingt, c'est le bout du monde. Mais celui qui peut s'acheter dix mille bicyclettes par seconde pendant dix mille ans, alors celui-là, il a peur ».

[3] Steven E. Jones, *Roberto Busa, S.J., and the Emergence of Humanities Computing. The Priest and the Punched Cards*, New York, Routledge, 2016.

[4] Theodor Mommsen et al. (éd.), *Inscriptiones Latinae antiquissimae ad C. Caesaris mortem*, vol. 1, Berlin, 1893.

[5] Thomas Kuhn, *The Structure of Scientific Revolutions*, Chicago, University of Chicago, 1962.

[6] Two key works about the historical method are: Johann Gustav Droysen, *Grundriss der Historik*, Leipzig, Veit, 1868; Marc Bloch, *Apologie pour l'histoire ou Métier d'historien*, Paris, Armand Colin, 1949.

## I. Towards a diplomatic corpus on a European scale

### *Qualifying and merging literacy*

First, I must say a few quick words about charters themselves. These documents are amongthe most frequently preserved text types of the Middle Ages, often kept as single sheets or in manuscript collections, called cartularies. They are found in almost every archive in Europe, and they cover a broad time period from roughly the Merovingian period to the 16th century - though with important typological variations. At least until the 13th century, these documents, first in Latin and then in vernacular languages, mainly refer to gifts, sales, exchanges or confirmations of land, goods and rights, most often made by a private actor to an ecclesiastical institution (abbey, priory, chapter, etc.)[8].

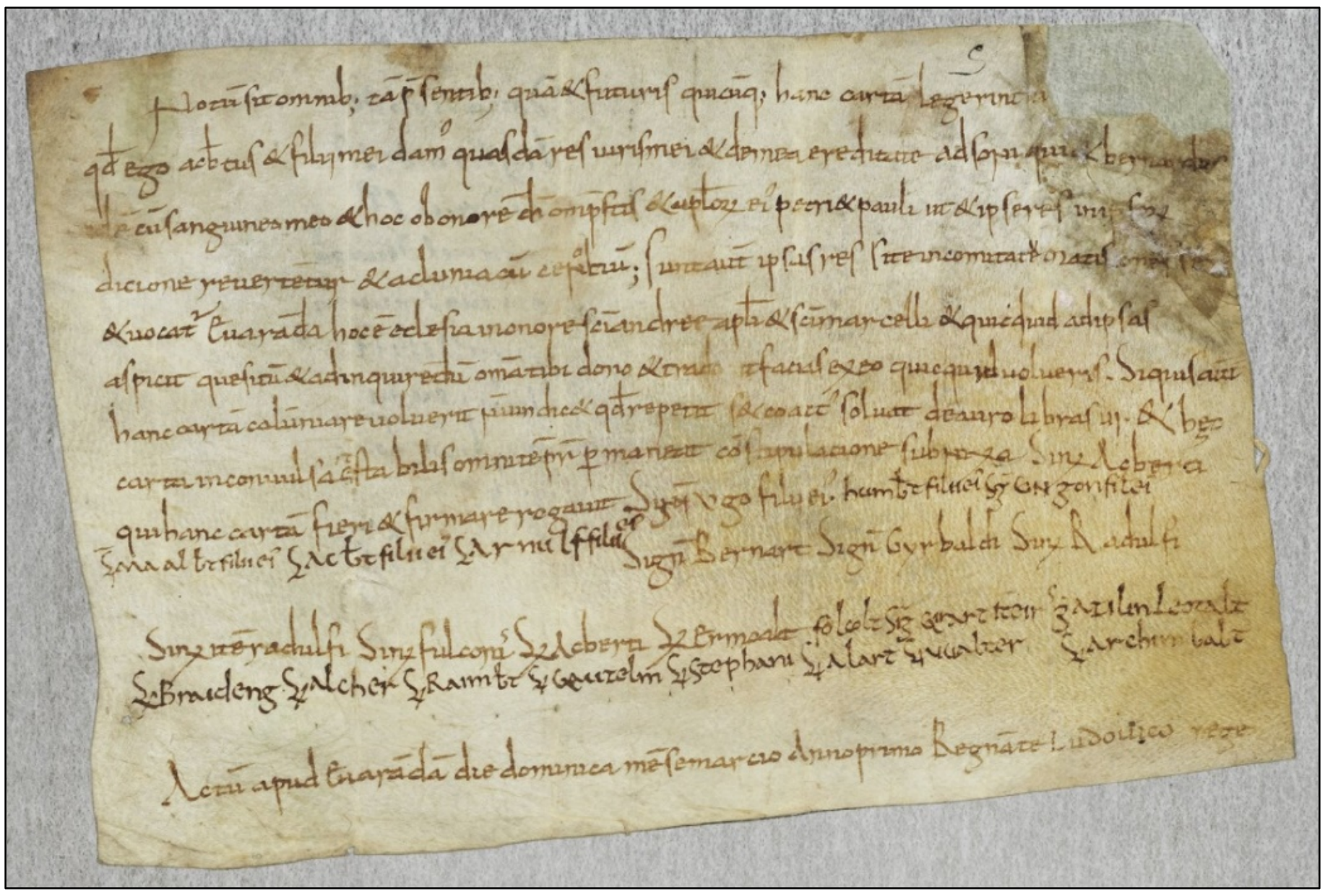

**Fig. 1:** A charter of Cluny: donation of Acbert to the abbey, March 937.

Paris, BnF, NAL 2154 (picture by Gallica).

[7] On this problem, see Alain Guerreau, *L'avenir d'un passé incertain. Quelle histoire du Moyen Âge au XXIe siècle ?*, Paris, Seuil, 2001.

[8] A great introduction to medieval charters and cartularies: Olivier Guyotjeannin, Jacques Pycke et Benoît-Michel Tock, *Diplomatique médiévale*, Turnhout, Brepols, 2006 (*L''atelier du médiéviste*, 2).

Begun in 2008-2009[9], the CEMA database now contains about 250,000 charters, or about 75 million words. The corpus has its roots in the contradictory situation already mentioned: on the one hand, there were many digitised diplomatic corpuses – I am thinking in particular of the *Chartae Galliae*, the CBMA, Scripta, the Cartulaire d'Île-de-France, the Deeds project, dMGH, or the *Codice Diplomatico della Lombardia Medievale*; on the other hand, these sites were not unified, and offered very disparate metadata, interfaces and formats. The first step in formalising the corpus was therefore to download and clean the available documents using dozens of scripts developed for the purpose. Another step consisted in formatting these files, after having performed a lemmatisation, so that they could be used for data and text mining.

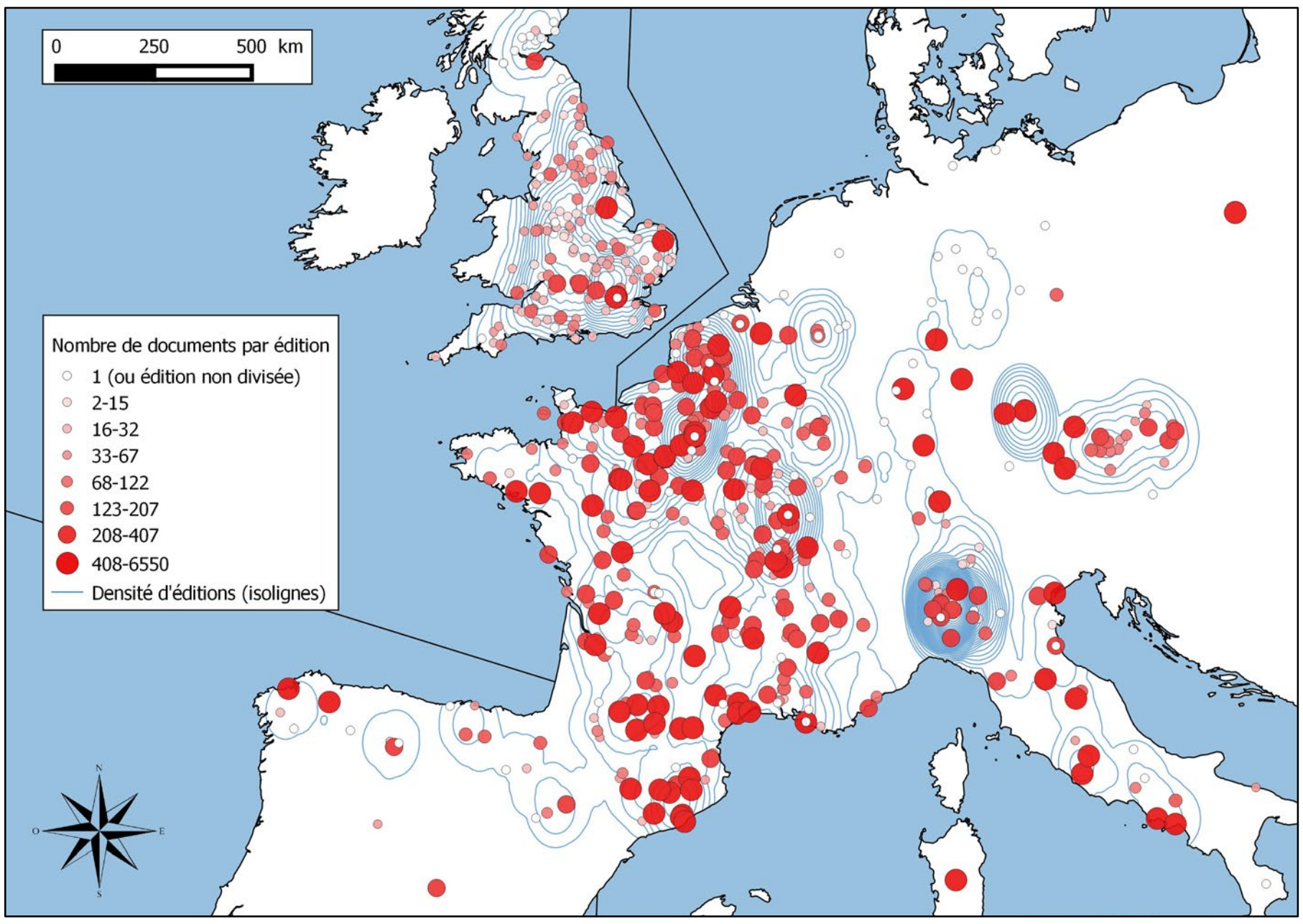

**Fig. 2:** Map of corpora available in the CEMA (circa 2015).

Why did I gather such a corpus? From my initial research, it appeared to me that there were very strong inequalities between the different European spaces in terms of written production. In order to study this phenomenon, various factorial analyses were applied to the number of charters per region and per half-century[10]. These investigations made it possible to produce different maps of the distribution of charters between the 7th and 14th centuries. You can see here a synthesis for the 10th-14th centuries. In order to make sure of the meaning to be given to this very clear distribution, I chose to map (so-called) Romanesque buildings as a parallel, by geolocating all the maps available in the Zodiaque collection[11], which show more than 8,000 buildings. The result is, as you can see, an almost perfect match between the areas where Romanesque buildings abound and those where charters from the 10th-11th or even 12th centuries are preserved *en masse*. This observation led me to question the process of documentary destruction, but above all also the limits of lexical analyses carried out on diplomatic databases.

[9] Nicolas Perreaux, *L'eau, l'écrit et la société. Étude statistique sur les champs sémantiques dans les bases de données*, Dijon, Université de Bourgogne 2010; Id., « L'eau, l'écrit et la société (IXe-XIIe siècle). Étude statistique sur les champs sémantiques dans les bases de données », *Bulletin du Centre d'études médiévales d'Auxerre*, n° 15, 2011, p. 439-449 (https://journals.openedition.org/cem/12062); Id., « Dynamique sociale et écriture documentaire (Cluny, Xe-XIIe siècle). Observations statistiques sur le champ sémantique de l'eau », in Dominique Iogna-Prat, Michel Lauwers, Florian Mazel, Isabelle Rosé (dir.), *Cluny. Les moines et la société au premier âge féodal*, Presses Universitaires de Rennes, Rennes, 2013, p. 111-128.

[10] Id., « L'écriture du monde (I). Les chartes et les édifices comme vecteurs de la dynamique sociale dans l'Europe médiévale (VIIe-milieu du XIVe siècle) », *Bulletin du Centre d'études médiévales d'Auxerre*, vol. 19.2, janvier 2016, https://cem.revues.org/14264; Id., « L'écriture du monde (II). L'écriture comme facteur de régionalisation et de spiritualisation du *mundus* : études lexicales et sémantiques », *Bulletin du Centre d'études médiévales d'Auxerre*, vol. 20.1, 2016, https://cem.revues.org/14452; Id., « De structures inconciliables ? Cartographie comparée des chartes et des édifices « romans » (Xe-XIIIe siècles) », in Marie-José Gasse-Grandjean, Laure Saligny (dir.)., *La géolocalisation des sources anciennes*, Dijon, 2016, https://cem.revues.org/13817 (*Bulletin du Centre d'études médiévales d'Auxerre*, Hors-série n° 9).

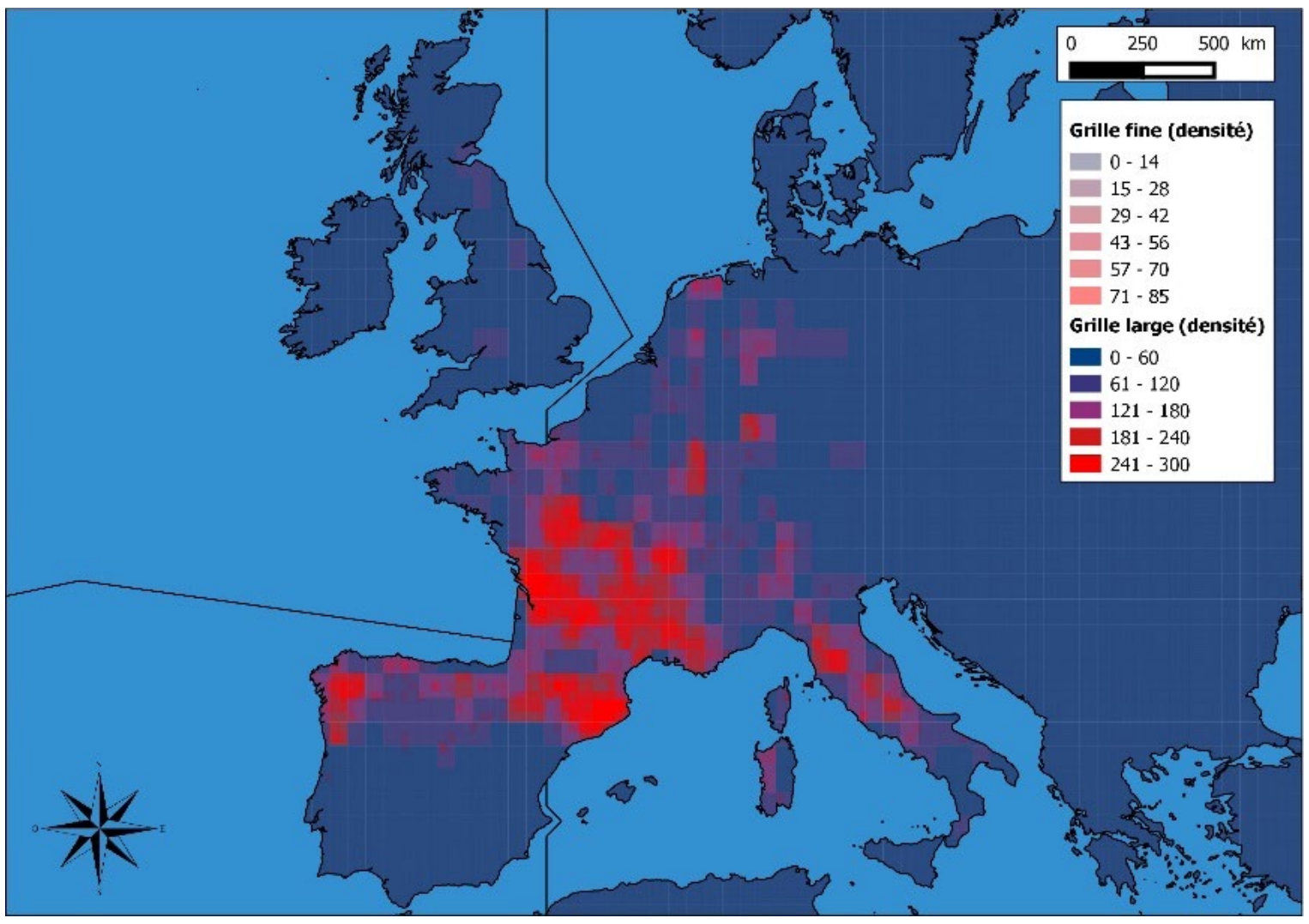

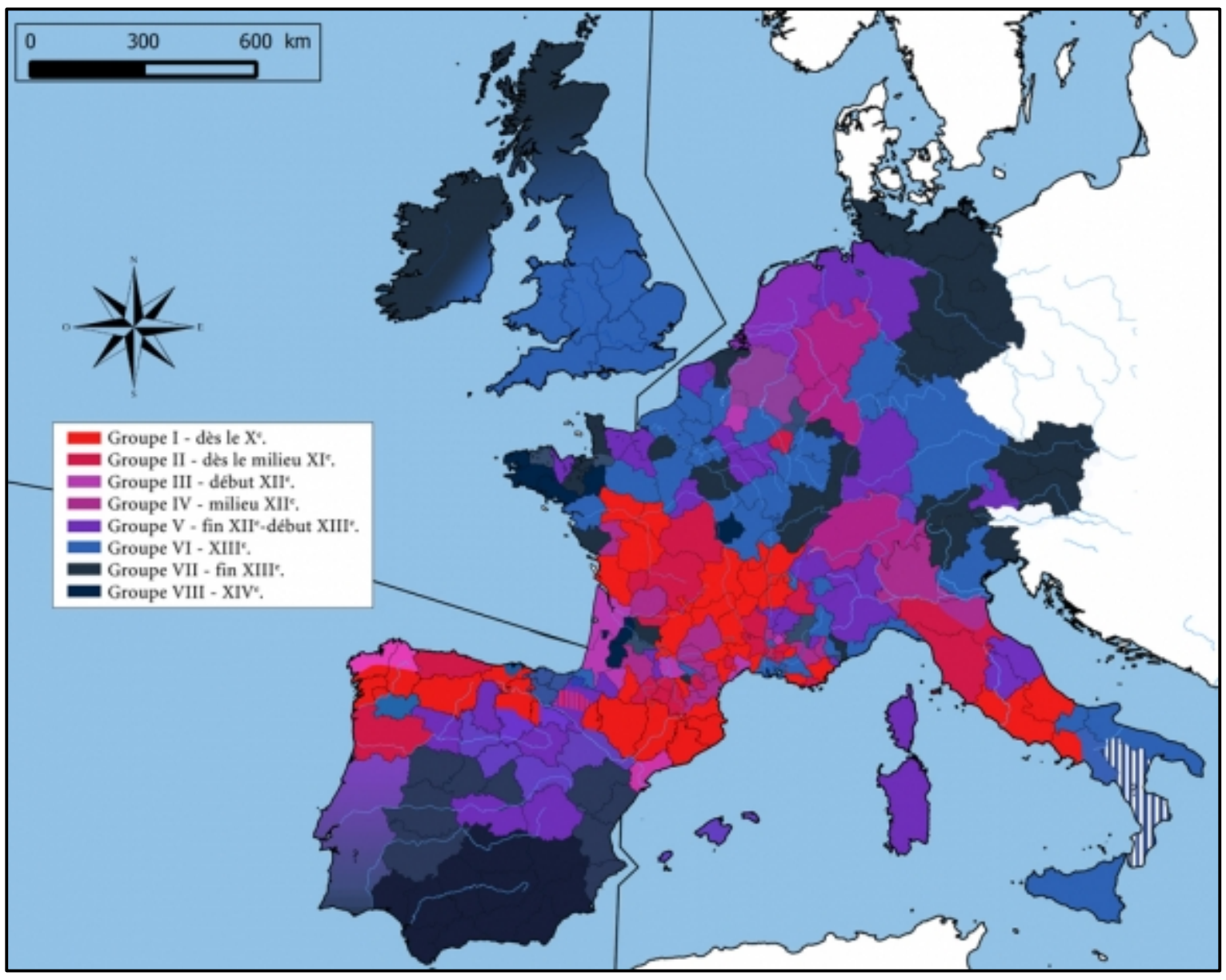

**Fig. 3a-b:** Distribution of charters and "Romanesque" buildings. (a, left). Mapping of more than 8,600 "Romanesque" buildings based on a geolocation of the Zodiaque collection (La nuit des temps). Density analysis via QGIS. (b, right). Mapping of the peak production of charters, by region, in Europe (10th-14thc.). Including 550,000 charters (2,000 editions). Processing by factorial analysis, clustering and QGIS. In red: the most dynamic regions in the 10th-11th centuries. We can see the almost perfect correlation between the "Romanesque" areas and those where charters were produced in the 10th-12th centuries.

## *Artificial intelligence, metadata and OCR*

How to develop the corpus in the near future? Two aspects must be distinguished here: texts and metadata. As far as metadata is concerned, thus far the choice has been made to retain all the information offered by the previous databases - from elements of authenticity to tradition, including mentions of seals. However, the latter are very disparate from one corpus to another. Artificial intelligence (AI) methods now make it possible to generate this metadata from a training corpus - this is called "learning" or even "deep learning". In 2011, the combination of different algorithms made it possible to distinguish bullae, royal and episcopal acts[12]. The aim is therefore to generate new metadata from existing ones quickly, in order to distinguish between document categories, types of action and languages, but also to date certain documents more precisely.

[11] For more information, see the Wikipedia article: https://fr.wikipedia.org/wiki/%C3%89ditions_Zodiaque

[12] Id., « Les documents pontificaux dans la masse (VIIe-XIIIe siècle). *Machine learning*, stylométrie et sémantique historique », in Olivier Guyotjeannin, Rolf Grosse, Laurent Morelle (dir.), *Les actes pontificaux. Un trésor à exploiter*, Berlin, De Gruyter, 2022 (*Abhandlungen der Akademie der Wissenschaften zu Göttingen. Neue Folge*, 51), forthcoming.

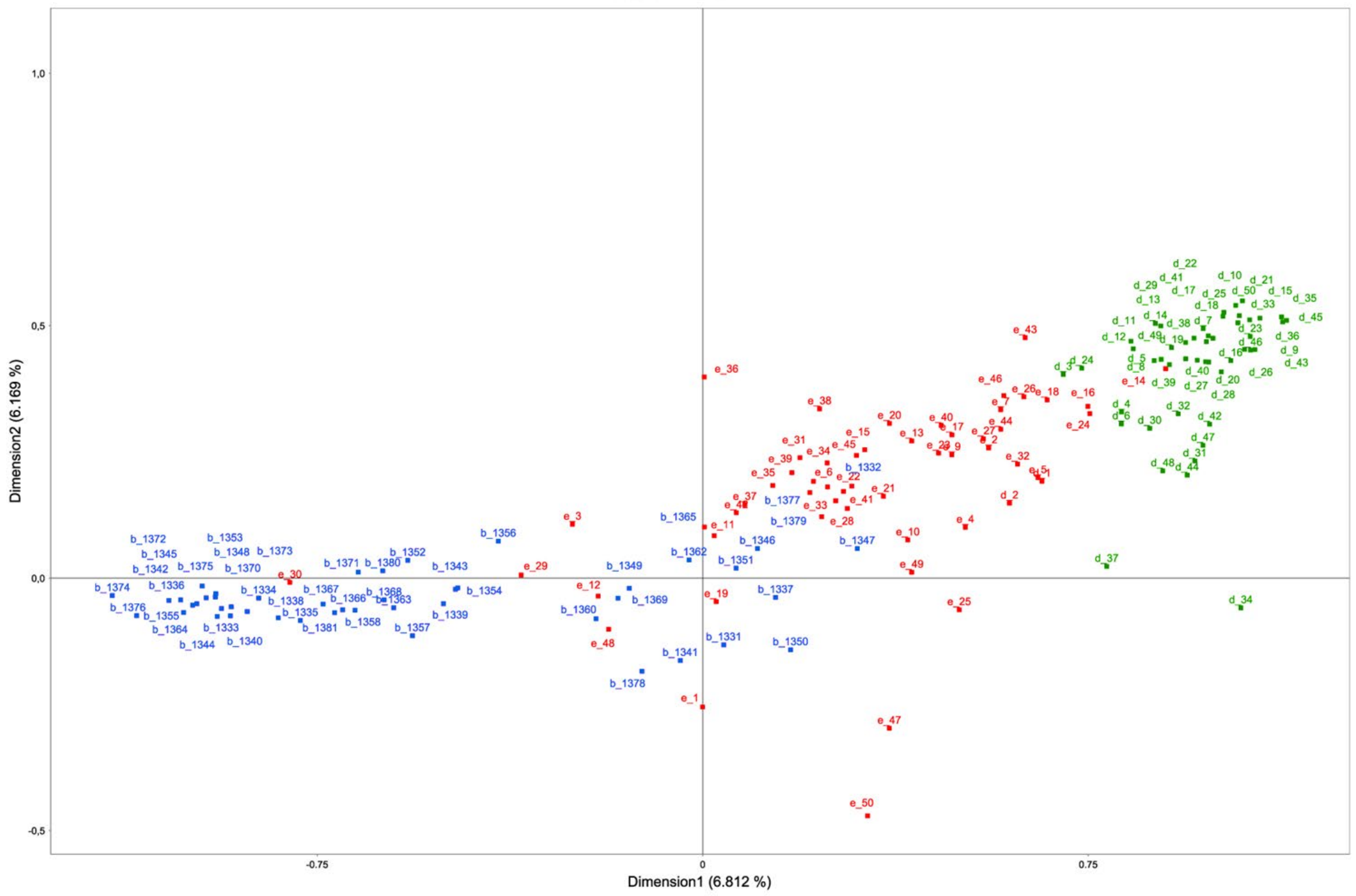

**Fig. 4:** Factorial analysis of a lexical matrix for different documentary typologies in the CEMA. In blue = bulls; in red = episcopal charters; in green = royal/imperial diplomas.

For example, as part of the colloquium "*Les actes pontificaux, un trésor à exploiter*", organised at the end of 2019 by Rolf Grosse, Olivier Guyotjeannin and Laurent Morelle at the Institut d'Histoire de l'Art (IHA, Paris), I was asked to give a paper on "pontifical texts as Big Data" (*Les actes pontificaux dans la masse*). However, in order to study them, we must first find these pontifical documents, which are not always identified as such in the CEMA. To do this, the computer must first be trained to distinguish between two basic categories: pontifical documents and... everything else, based on the lexicon of the acts. It is therefore necessary to start from corpora that already list bulls, such as the CBMA, *Diplomata Belgica* or the Artem. Once this training has been carried out, we obtain a model capable of classifying the acts. In this case, the result was the identification of a corpus of 26,000 pontifical acts, which opens the door to comparative investigations. By generalising these methods, the hoped-for result is twofold: 1) a corpus that can finally be exploited in much greater detail; 2) new knowledge about the writing of charters from this or that chancery or institution.

The other semantic field for the extension of the CEMA is now well known: optical character recognition. The study of diplomatic prides itself on its advances in handwritten text recognition (HTR), which is a possible way for future corpora, so the CEMA could already benefit from an extension based on the OCR of existing editions in "image mode"[13]. By creating an adapted processing chain, it is safe to say that hundreds of thousands of additional acts could be added to the corpus. This is why, in parallel with the online publication of the corpus, by the end of the academic year, in partnership with the LaMOP, the École nationale des Chartes, the IRHT-Telma, the MGH, the CSIC and other national and international institutions/teams, we plan to publish a bibliography of European diplomatic editions, as well as several thousand diplomatic editions in image mode, that I patiently collected over many years[14].

[13] This perspective was presented in Zürich, in 2018, at the invitation of Simon Teuscher, whom I thank warmly. I discussed it at different stages afterwards, in particular during the Workshop *Humanités numériques. De nouveaux outils pour le médiéviste* (dir. Paul Bertrand, Nicolas Ruffini and Sébastien de Valeriola), at Namur and Louvain.

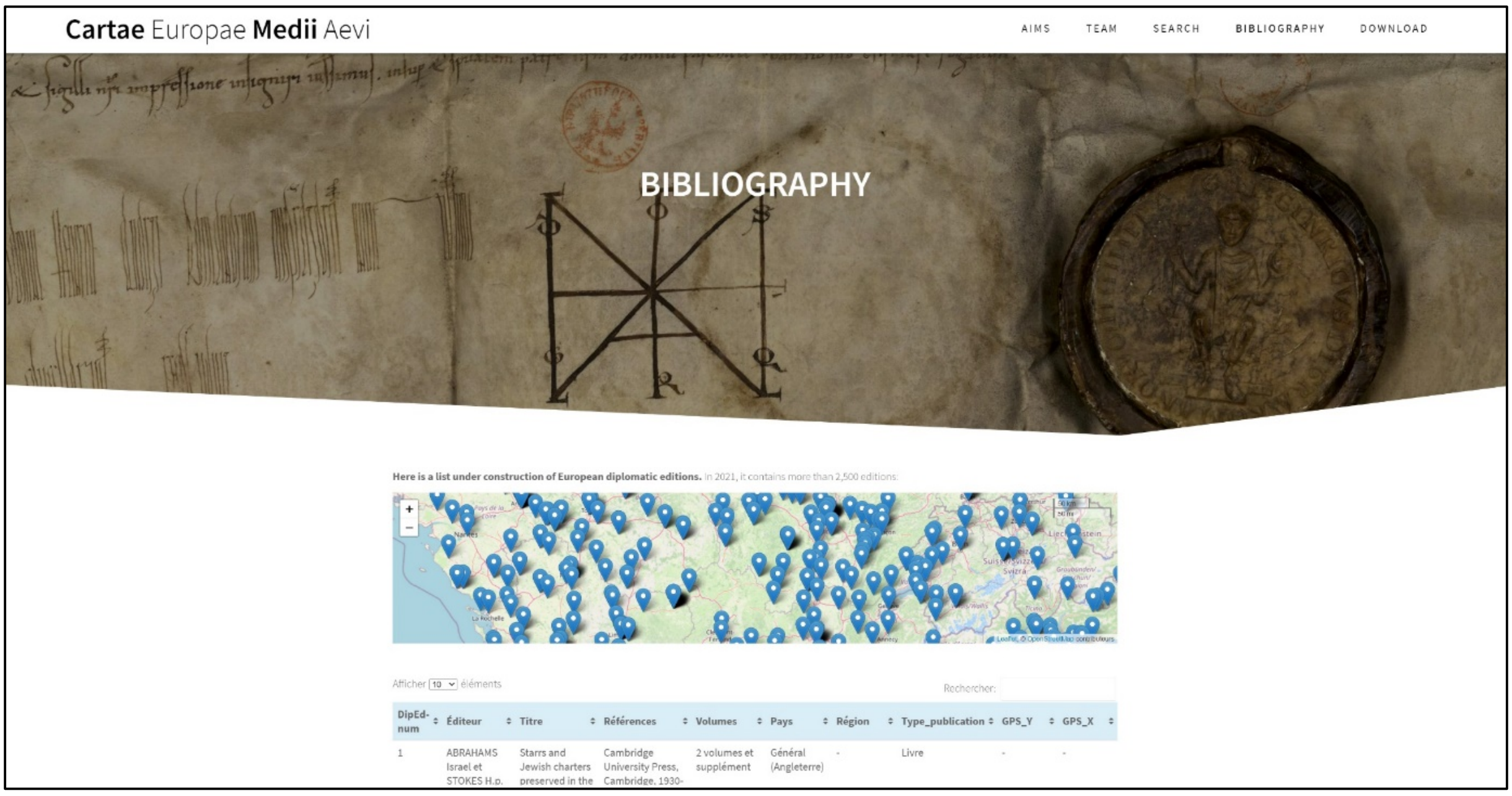

**Fig. 5:** Screenshot of the CEMA website.

## *Measuring lexical richness*

How then to go further and measure the richness of texts included in the CEMA[15]? This is a recurrent difficulty in diplomatic, which raises various questions in relation to the forms, lexical and calligraphic knowledge of the scribes. It is an aporia that many historians face, but which can be solved thanks to the contribution of mathematicians and linguists. The process is relatively simple: the algorithm calculates the number of unique lemmas in a text or corpus, starting from the first word and going to the end. Initially, all words encountered are new, but gradually the words the program inspects have already been seen. So the curve bends. Thus, the faster the curve rises and stays high, the richer the lexicon. Better still, as this richness can be compared in a package of 500 words, 1,000 words or 10,000 words, this approach makes it possible to compare corpora of different sizes.

[14] https://lamop.pantheonsorbonne.fr/axes-recherche/livres-textes-langages/cema. My thesis is now online in open access: N. Perreaux, *L'écriture du monde. Dynamique, perception, catégorisation du mundus au moyen âge (VIIe-XIII*[e] *siècles). Recherches à partir de bases de données numérisées*, Dijon, 2014, https://halshs.archives-ouvertes.fr/tel-03084322

[15] Various tools used in the following pages to explore the lexical content of documents are taken from the Cooc function library - developed by Alain Guerreau in the framework of the ANR Omnia.

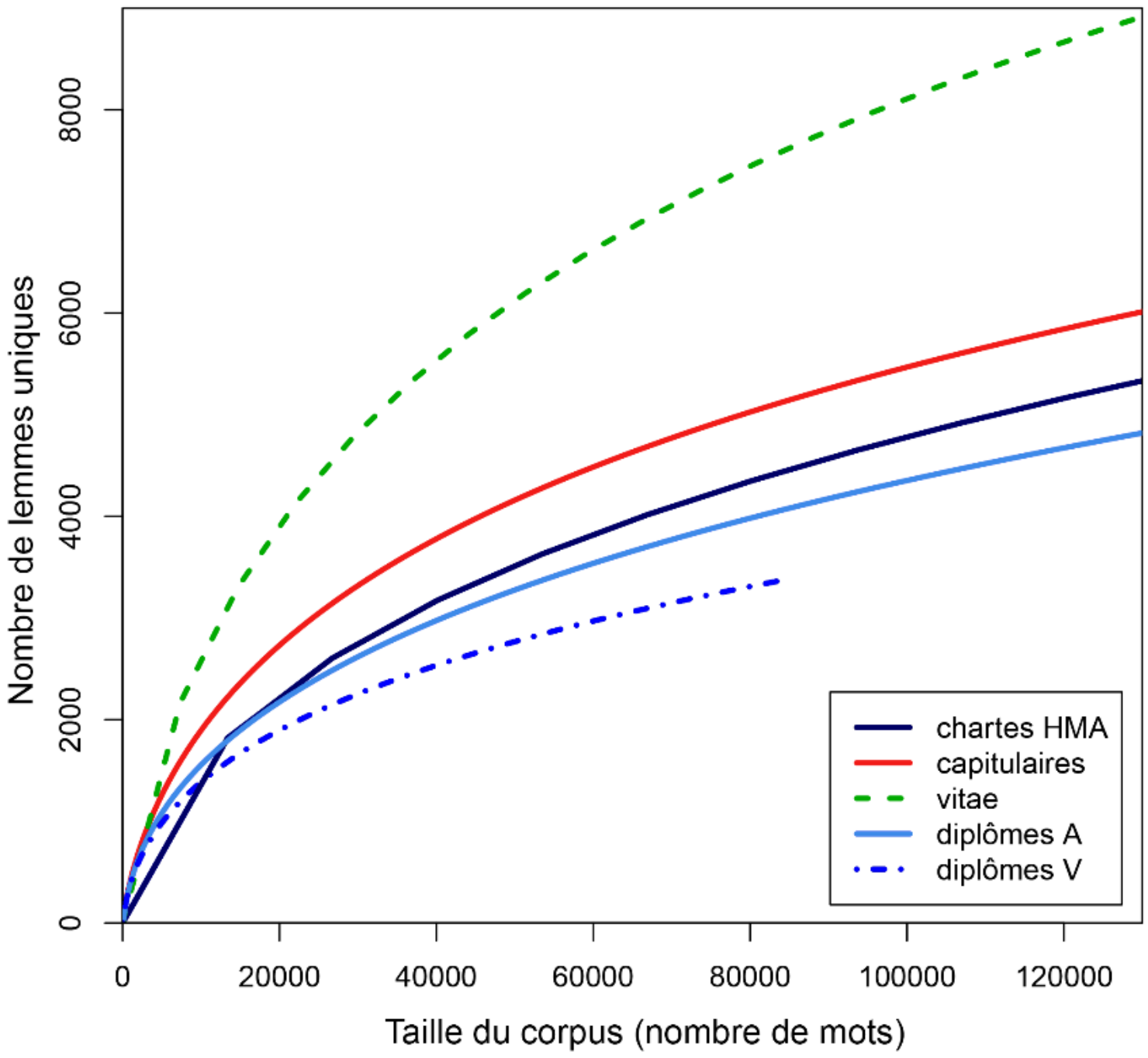

**Fig. 6:** Measurement of the lexical richness in some corpora of the Early Middle Ages: charters (all types), capitularies, *vitae*, diplomas. Method: ZipfR (S. Evert) and Cooc (A. Guerreau).

When working on papal documents[16], I wondered whether historians knew when the lexicon of bulls was richest. The results of a small survey on Twitter, presented here, pointed to the twelfth-thirteenth centuries... but mostly showed a great deal of hesitation. The survey shows some relatively unexpected results[17]. As you can see on the left, the lexical richness of the 7th-9th century records is indeed significantly higher than that of the 10th-12th centuries on the right. These results, confirmed in a century-by-century analysis, contradict several historiographical intuitions - in particular that the use of pontifical formularies would have strongly standardised their writing in the early Middle Ages.

[16] Id., « Les documents pontificaux dans la masse […] », *op.cit.*

[17] My question was: « In your opinion, the lexicon of pontifical documents is richer in a) the 7th-9th century; b) the 10th-11th century; the 12th-13th century; no idea ». Of the 28 votes obtained, 36% were for the 12th-13th centuries, 21% for the 10th-11th centuries, 18% for the 7th-9th centuries, and 25% said they had no hypotheses.

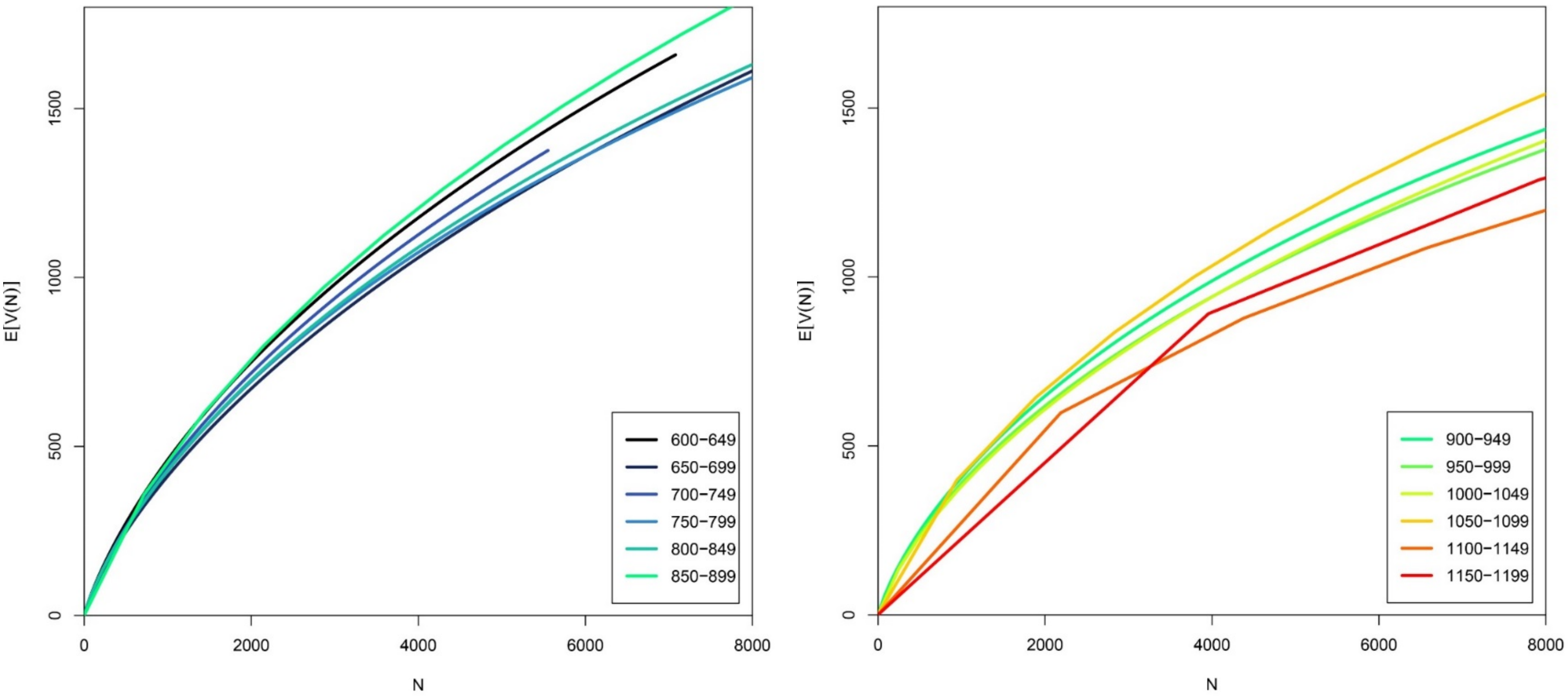

**Fig. 7a-b:** Richness of the lexicon of pontifical documents (lemmas), by half-century. On the left, for the period 600-899. On the right, for the period 900-1199. Method: ZipfR (S. Evert) and Cooc (A. Guerreau).

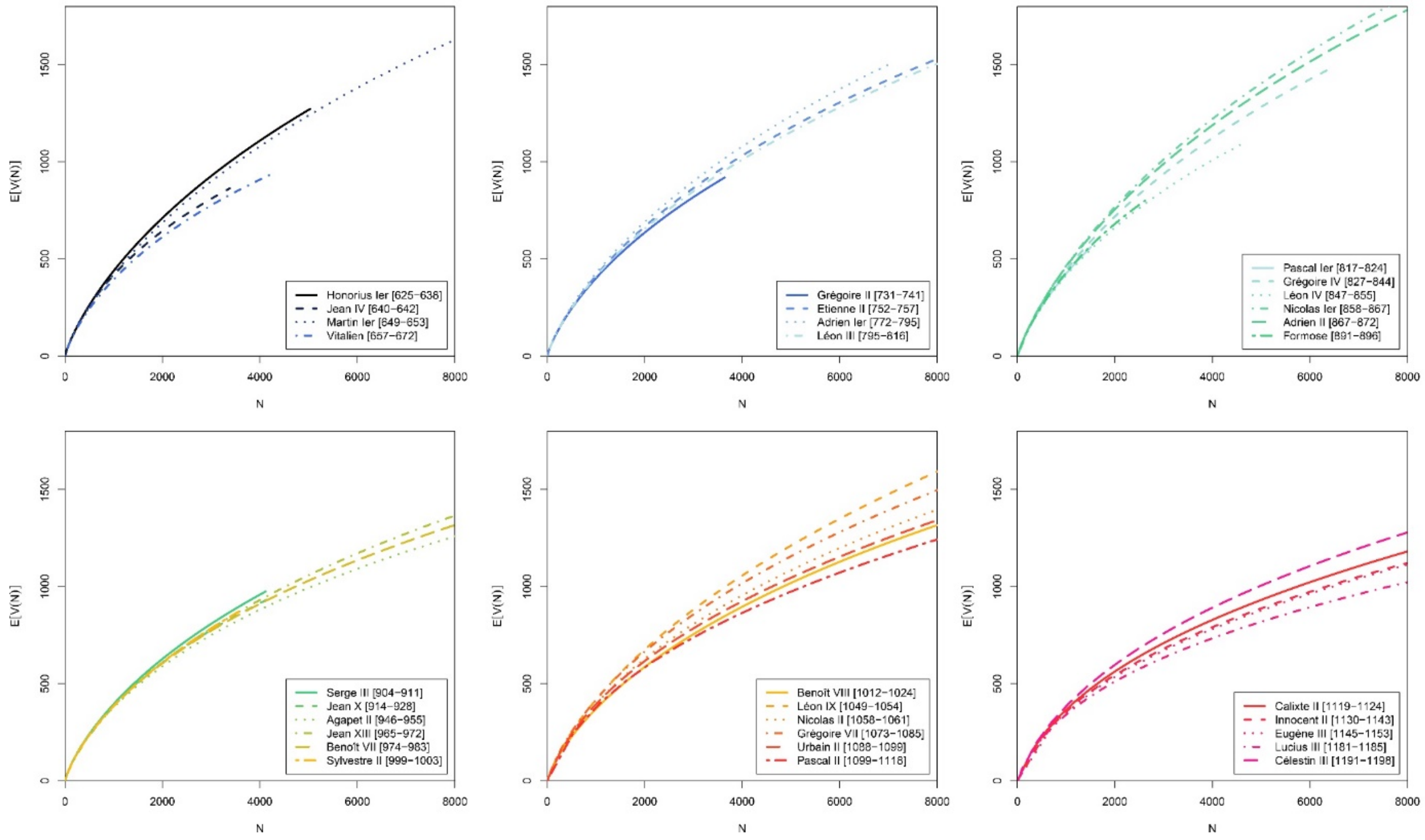

**Fig. 8a-f:** Richness of the lexicon of pontifical documents (lemmas), for various pontificates (7$^{th}$-12$^{th}$ centuries). Method: ZipfR (S. Evert) and Cooc (A. Guerreau).

Many other examples could be used here, but I will choose just one. This graph compares the wealth of Merovingian and Carolingian capitularies and diplomas[18]. Two trends can be observed here: the capitularies are richer overall than the diplomas. But we also note that the writing of the two corpora tends to become richer between the second half of the 8th century and the second half of the 9th century, which is quite amusing. The diplomas of the successors of Louis the Pious are more varied than the capitularies of before 840. If these experiments were generalised, they would provide a better understanding of the history of lexical variability, and thus of the standardisation of medieval institutions.

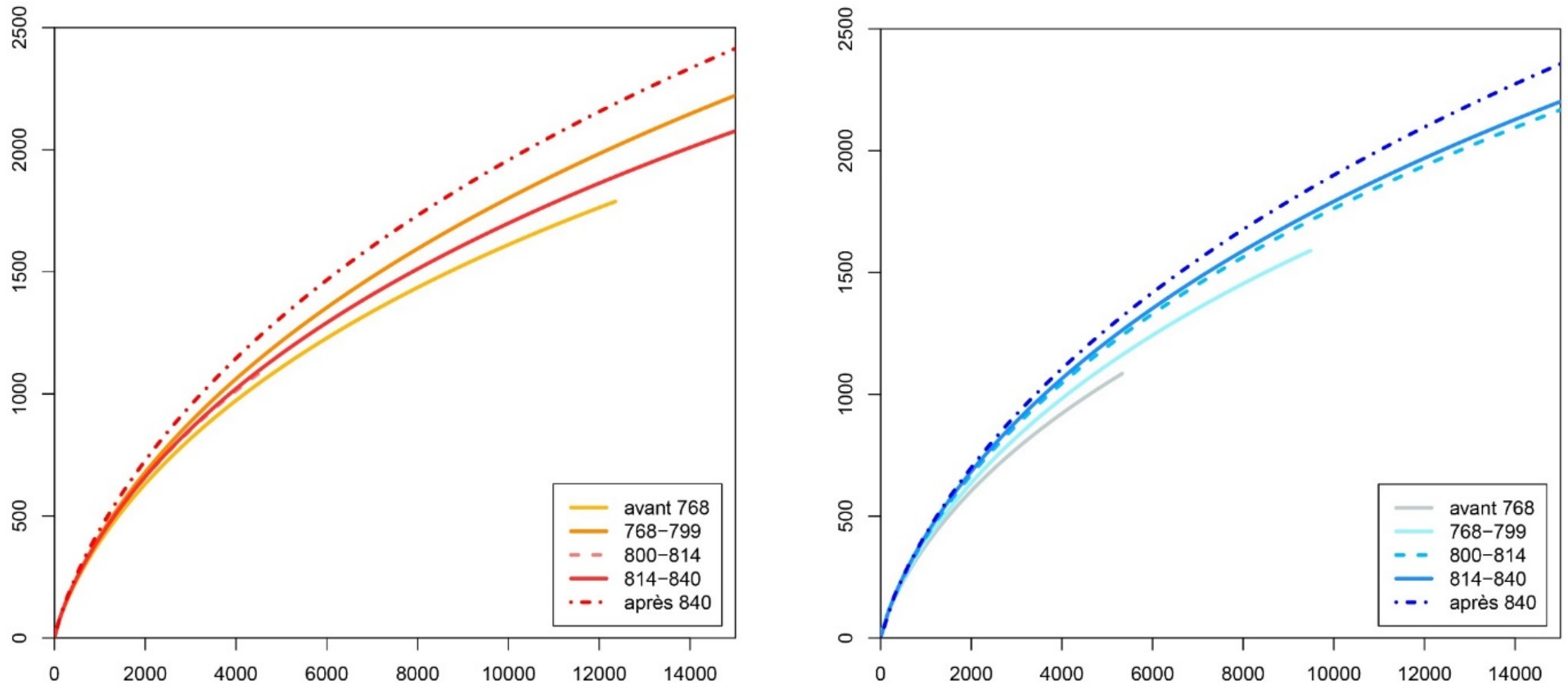

**Fig. 9a-b:** Comparative richness of the lexicon (lemmas) of Merovingian/Carolingian capitularies (left) and diplomas (right) in different time periods.

## II. Exploring time and space

### *Medieval chrono-geography*

Another field in which the CEMA could produce interesting results is that of the chrono-geography of medieval phenomena – a theme which is related to the more global one of lexical geography and dialectology. For example, we still know very little about the evolution of mills, the vocabulary of kinship, or spatial entities. Yet the periodisation of these elements plays a decisive role in our understanding of medieval dynamics. The great syntheses of medieval Europe – those of Georges Duby, Léopold Génicot, Robert Fossier but also Chris Wickham or Florian Mazel[19] – mainly use regional monographs in order to establish their chronologies. This is both logical and necessary, but the result is a complex patchwork, as the authors of the monographs themselves focus on different points: sometimes the infrastructure, sometimes the written word, kinship, coins, or even aristocratic.

[18] Id., « Langue des capitulaires et langue des chartes : richesses, circulations, spécificités », in Bernhard Jussen, Karl Ubl (dir.), *La langue du droit. Sémantique historique et capitulaires carolingiens*, Frankfurt, 2021 (forthcoming).

[19] Georges Duby, *L'économie rurale et la vie des campagnes dans l'Occident médiéval (France, Angleterre, Empire, IXe-XVe siècles)*, 2 volumes, Paris, Aubier, 1962; Léopold Génicot, *Les lignes de faîte du Moyen Age*, Tournai, Casterman, 1969; Robert Fossier, *Enfance de l'Europe, Xe-XIIe siècle*, 2 volumes, Paris, Presses universitaires de France, 1982; Chris Wickham, *Framing the Early Middle Ages. Europe and the Mediterranean, 400-800*, Cambridge, Cambridge University Press, 2005; Florian Mazel, *Féodalités, 888-1180*, Paris, Belin, 2010.

An enlightening case is the remarkable *Enfance de l'Europe* by Robert Fossier[20]. In this masterly work, the author presents thematic assessments by using regional comparisons. He successively discusses the production of charters, demography, the creation of new towns, land clearing, Motte-and-bailey castle, levies, etc. In total, almost 26 typo-chronological criteria of medieval dynamics are examined. If we formalise these passages, we see that they compare 67 zones (at least), sometimes vast (such as England or France), sometimes smaller (Berry, the Chartres region, Paris, etc.). But we also see that it is almost never the same areas that are present in the comparisons. This is logical, as Robert Fossier bases his comparisons on monographs. But it is also problematic, as the examination of all the criteria in all the areas surveyed would result in a table of more than 1,700 combinations.

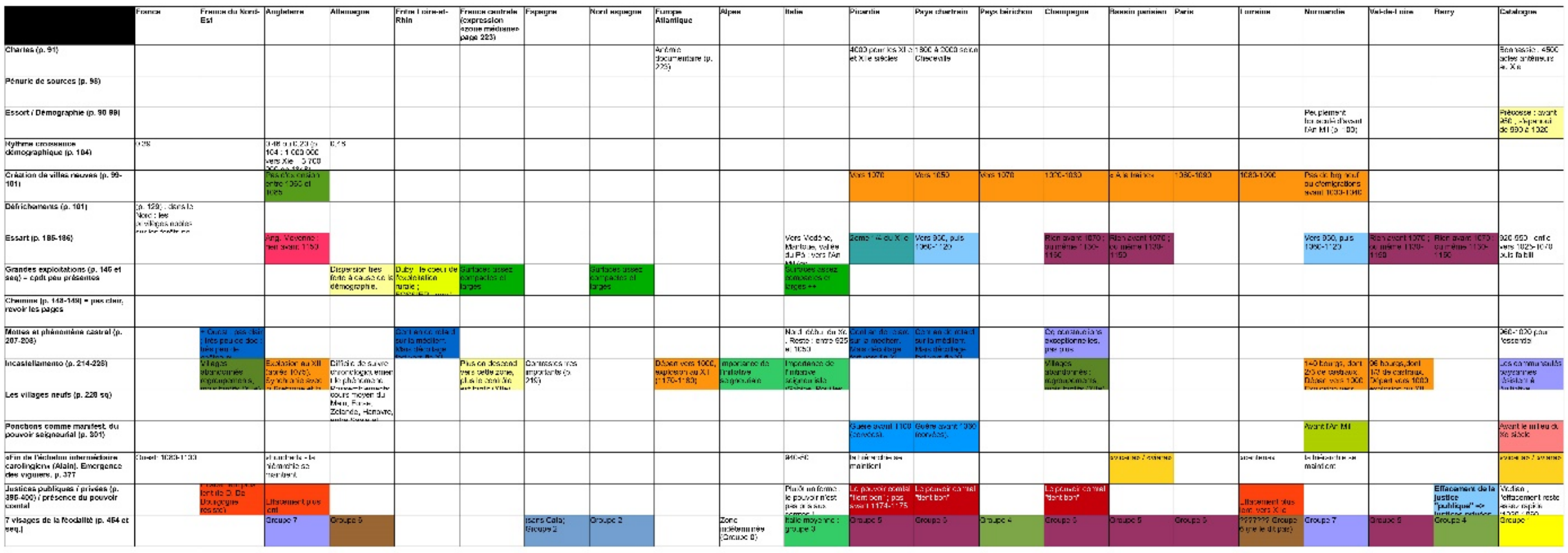

**Fig. 10:** Extract from the table of criteria/areas/chronologies in Robert Fossier's *Enfance de l'Europe* (2 volumes, 1982). 67 zones are presented (at least); 26 typo-chronological criteria (at least); *i.e.* 1,742 possible combinations, from various monographs.

## *Temporalities*

The CEMA make it possible to carry out chronological interrogations, by varying the scales of analysis. Four examples are given here: a) burials, b) mills, c) spatial structures and d) storage sites. In all cases, a simple method has been used: it consists of dividing the corpus into chronological slices containing an equal number of words. Medieval corpora vary in size from one period to another; the aim is therefore to measure the evolution of words by taking into account the variability of the corpus itself. However, to collect, for example, 1 million words, it sometimes takes a century and sometimes ten years. Once these equivalent slices have been created, it is sufficient to count the occurrences of the terms in order to observe their evolution. In this case, I have again used an algorithm coded by Alain Guerreau, in the R:Cooc library of functions.

The first example concerns medieval burials. In this semantic field, the charters initially indicate a high number of isolated burials with the term *sepultura*[21]. Then, from the 11th century onwards, the terms *cimeterium* and atrium are used in large numbers. This movement coincides with the fall in the number of mentions of *sepultura*. This lexical evolution confirms several hypotheses of Michel Lauwers, and raises various questions[22]: are cemeteries newly created, or are they taken over, within the framework of the capture of rights over and around burials by the Church?

[20] Robert Fossier, *Enfance de l'Europe […]*, op.cit.

[21] Nicolas Perreaux, « Les mots de la sépulture dans l'Europe médiévale (VIIIe-fin du XIIIe siècle) : observations complémentaires à partir des corpus numérisés », in Michel Lauwers, Aurélie Zemour (dir.), *Qu'est-ce qu'une sépulture ?*

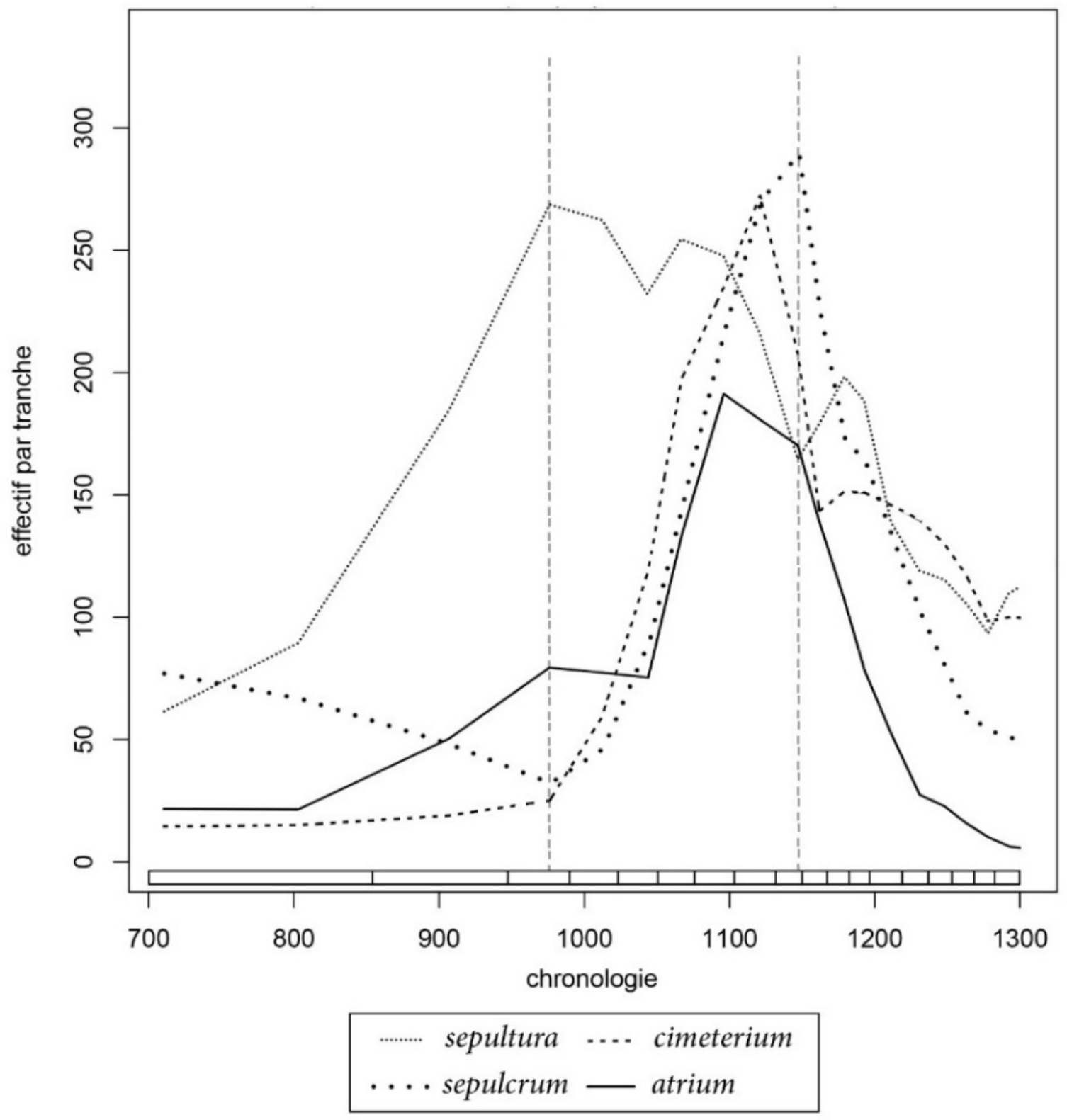

**Fig. 11:** Chronology of grave mentions (*sepultura / cimeterium / sepulcrum / atrium*) in the CEMA, 8th-13th centuries. Cooc function for R-CWB/Rcqp.

In the case of *molendinum* and *farinaria* (the two main terms used to designate mills), the analysis allows for a dialogue with historiography, which since Marc Bloch has oscillated between two hypotheses: firstly, the one that puts forward the diffusion of mills during the 11th-13th centuries; secondly, the one that supports the existence of these infrastructures as early as the early Middle Ages. Here, the graph shows that a middle way can be defended: mills do exist in the acts of the early Middle Ages; but this does not prevent their number from rising in charters from the middle of the 9th century, and more particularly in the 11th and 12th centuries. Again, this is probably for both material and ideological reasons.

*Humanités et systèmes funéraires de la préhistoire à nos jours. Actes des XXXVI^e^ rencontres internationales d'archéologie et d'histoire d'Antibes, 13-15 octobre 2015*, Antibes, Éditions APDCA, 2016, p. 113-121.

22 Michel Lauwers, *Naissance du cimetière : lieux sacrés et terre des morts dans l'Occident médiéval*, Paris, Aubier, 2005.

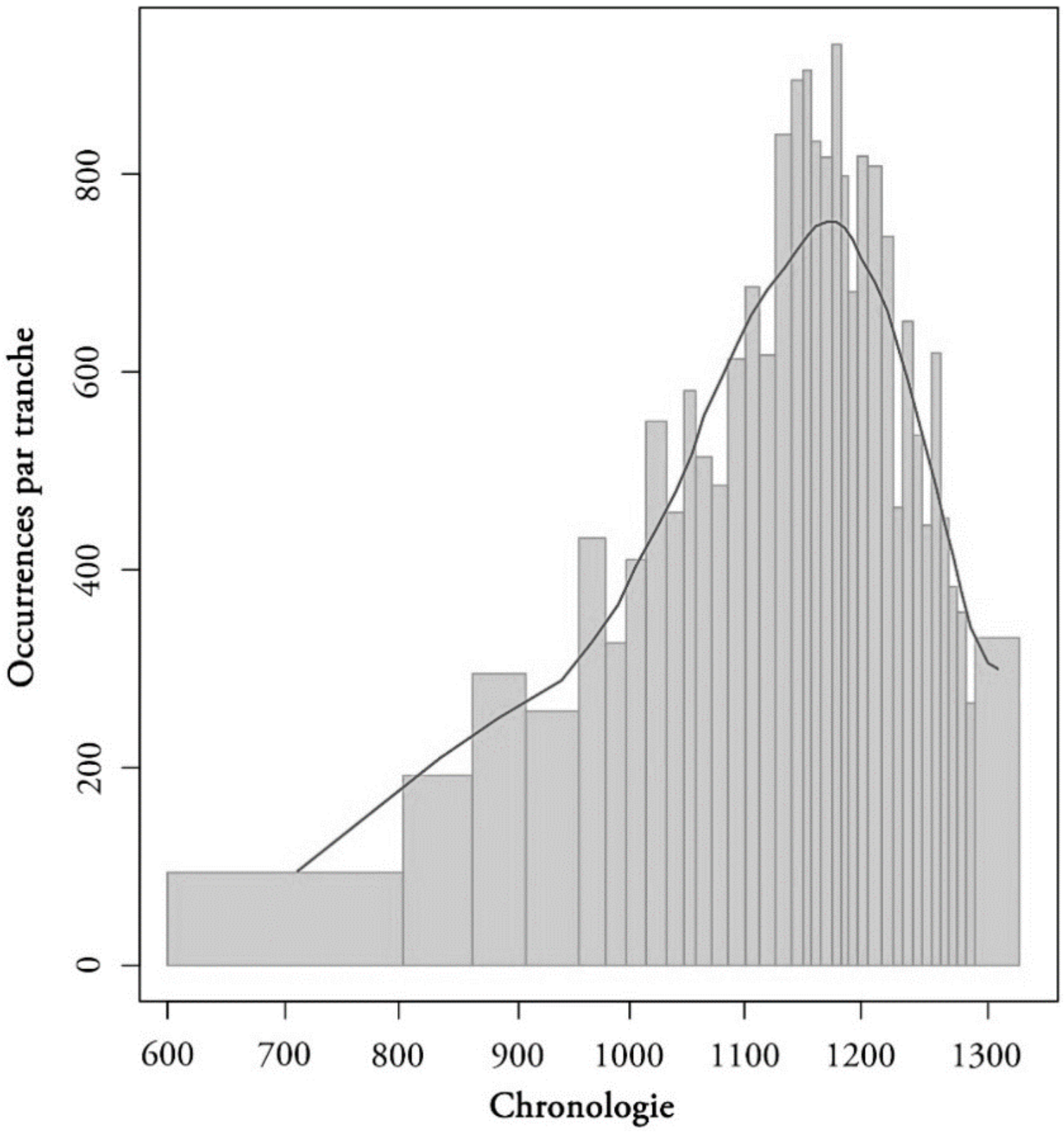

**Fig. 12:** Chronology of mill mentions (*molendinum / farinaria*) in the CEMA, 7th-13th centuries. 21,392 occurrences are included in this analysis. Cooc function for R-CWB/Rcqp.

In parallel, I have recently carried out an investigation into the evolution of spatial structures or entities linked to seigneurial domination in the CEMA[23]. Starting from a list of forty terms, including *villa*, *pagus*, *ager*, *comitatus*, *fundus*, *provincia*, *parochia*, *diocesis*, etc., it was possible to distinguish the elements that played an important role, and to propose a scheme of the evolution of these structures. The study showed that the triad of *villa*, *ager* and *pagus* played a central role in the location of lands/goods/sites from the 7th to the 8th centuries. In the middle of the 9th and especially 10th centuries, this pattern was enriched by the arrival of *comitatus* and *vicaria*, which did not replace but supplemented the previous triad. However, from the middle of the 11th century onwards, almost all of these spatial entities disappeared, to the benefit first of all of *parochia* and then of *diocesis*. In this new framework, *feodum* is integrated, but also *villa*, which is endowed with a completely new meaning. This chronological analysis, combined with semantic investigations, has thus made it possible to show that there is a transition, in the 11th-12th centuries, from a system in which spatial entities were essentially linked to the functions of the lay aristocracy, to a system in which only the *ecclesia* controlled the most comprehensive structures.

[23] Nicolas Perreaux, « Des « seigneuries » laïques aux territoires ecclésiaux ? Dynamique du processus de spatialisation dans les actes diplomatiques numérisés (VIIe-XIIIe siècles) », in Tristan Martine, Jessika Nowak, Jens Schneider (dir.), *Espaces ecclésiastiques et seigneuries laïques. Définitions, modèles et conflits en zones d'interface (IXe-XIIIe siècle)*, Paris, Presses Universitaires de la Sorbonne, 2021, p. 55-77.

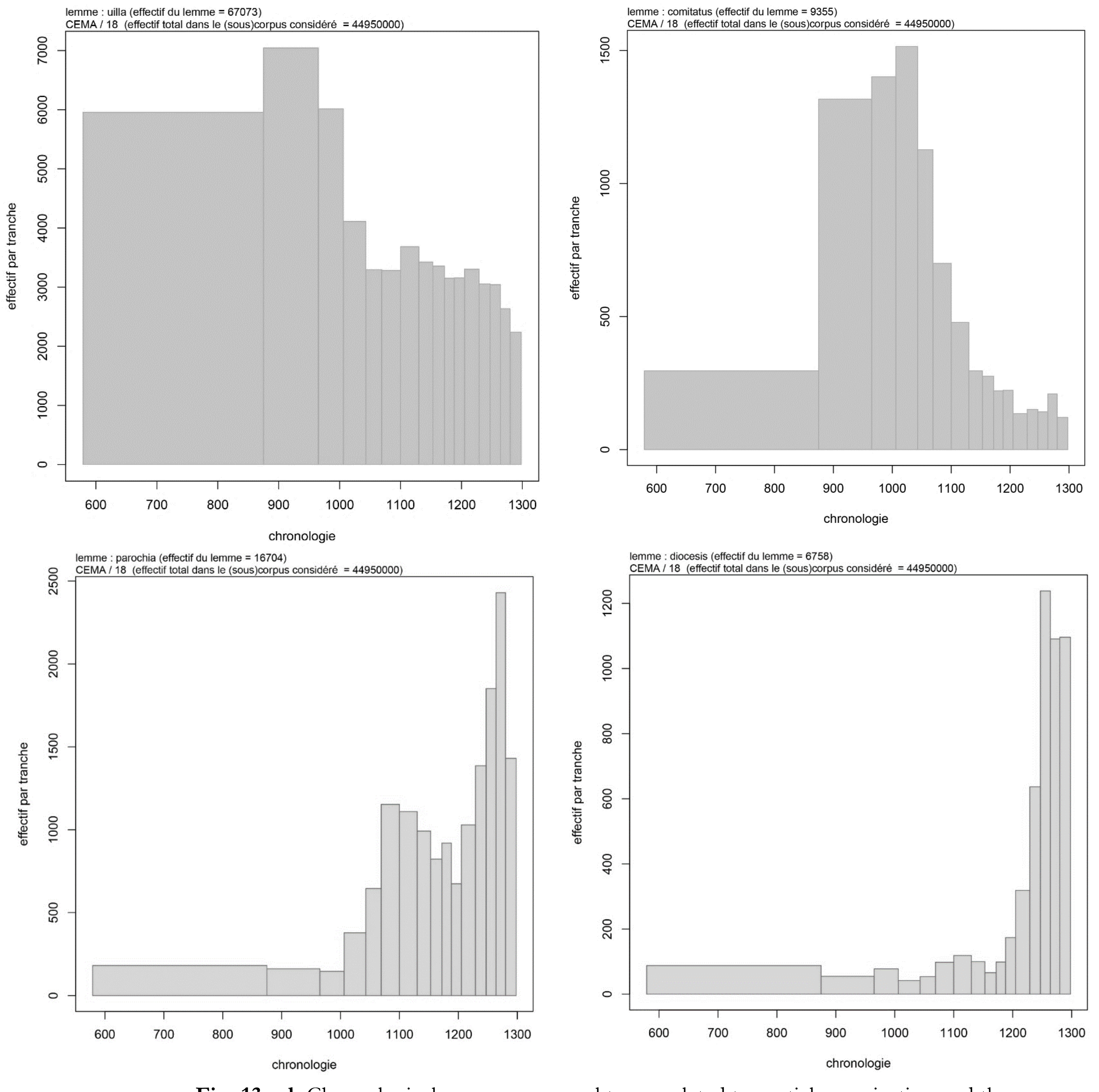

**Fig. 13a-d:** Chronological survey on several terms related to spatial organization and the dominium. (a) *villa*; (b) *comitatus*; (c) *parochia*; (d) *diocesis*. An analysis of the graphs makes it possible to define a model of the evolution of these structures.

The last example selected concerns storage places, on which I worked in the framework of the 2018 Flaran meetings[24]. The graph shows the evolution of the cumulative occurrences of *grangia*, *granarium*, *granica*, *horreum*, *spicarium* and *granea*. Several things are apparent. The first is the rarity of these structures in the acts of the early Middle Ages. This observation contrasts sharply with what is observed by archaeologists concerning silos, which are increasingly frequently found both in the north and in the south. Without setting texts and excavations up as contradictory, the proposed hypothesis is that this difference is attributable to the changing nature of the social relationships visible in the charters. In addition to the creation of new storage structures, what changes is the attitude to the taking of goods: while the acts of the early Middle Ages focus largely on the practice of giving, in the 11th century various questions concerning the taking of goods become more widespread. From then on, workers had to bring a share of their production to the land-holders, who built ad hoc granaries for this purpose, made of wood or stone, outside the production sites.

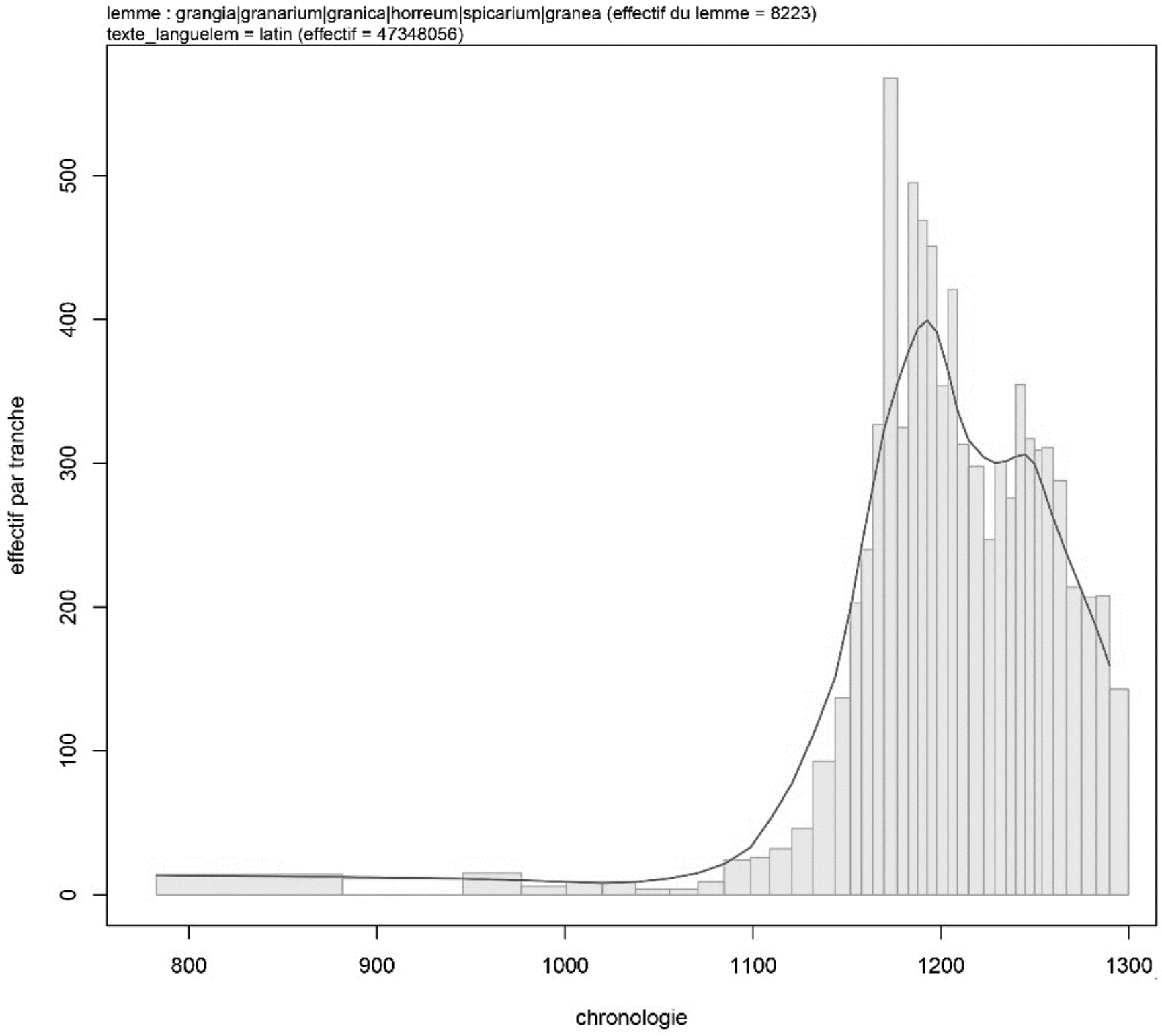

**Fig. 14:** Chronology of cumulative occurrences for *grangia*, *granarium*, *granica*, *horreum*, *spicarium* and *granea* in the CEMA (late 8th-13th c.).

## *Aires de scripturalité* and the regional writings of charters

However, it would be right to point out that these trends varied from one region to another. Medieval Europe is characterised, as we know, by strong inter- and intra-regional disparities. This phenomenon is well known to linguists, archaeologists, art historians and textual historians. In

[24] Id., « Les lieux de stockage dans les textes diplomatiques (VIIe-XIIIe siècles). Enquête lexicale, sémantique et numérique », in Michel Lauwers, Laurent Schneider (dir.), *Mises en réserve : production, accumulation et redistribution des céréales dans l'Occident médiéval et moderne*, Flaran, PUM, 2021 (forthcoming).

2010, Florian Mazel still insisted on "the great diversity of regional and local situations"[25]. This diversity manifests itself in different ways in the acts, first of all through lexical regionalisms, which are little studied for Latin.

An initial approach may consist of spot surveys, initially cartographic, of this or that term or expression. The example chosen in the first instance is that of *finagium*[26]. The term is quite common in rural history, where the "finage" makes a regular appearance. However, this historiographical usage masks the fact that the Latin lemma has a specific distribution in diplomatic texts, as you can see - with a strong concentration in the east of present-day France.

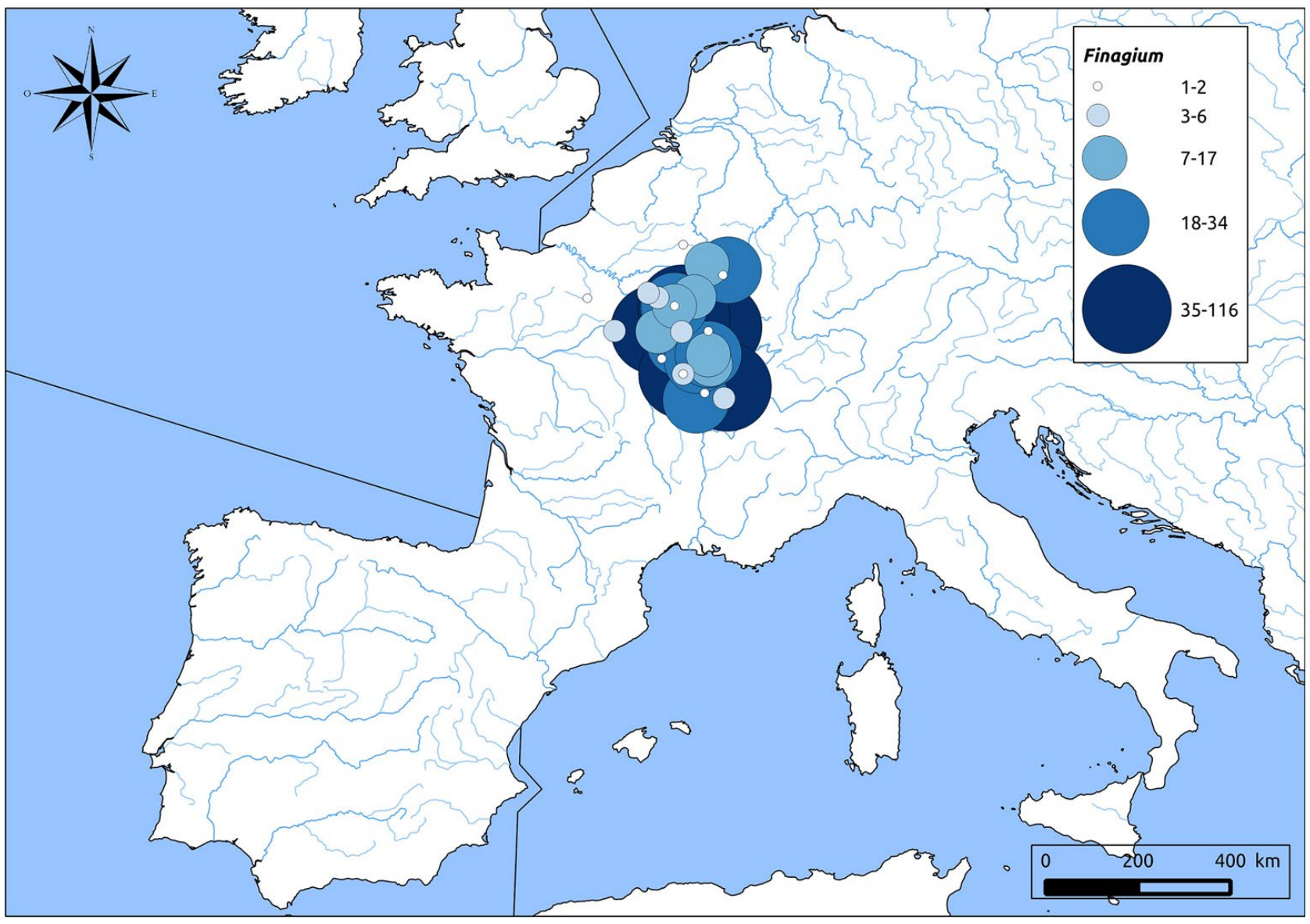

**Fig. 15:** Geographical distribution of *finagium* occurrences in the CEMA.

These experiments could of course be generalised. In my thesis, I identified some of these regionalisms, with for example *molina*, the formula *pomiferis and inpomiferis* or the lemma *tellus* - which frequently replaces *terra*, but almost exclusively in southern Insular documents.

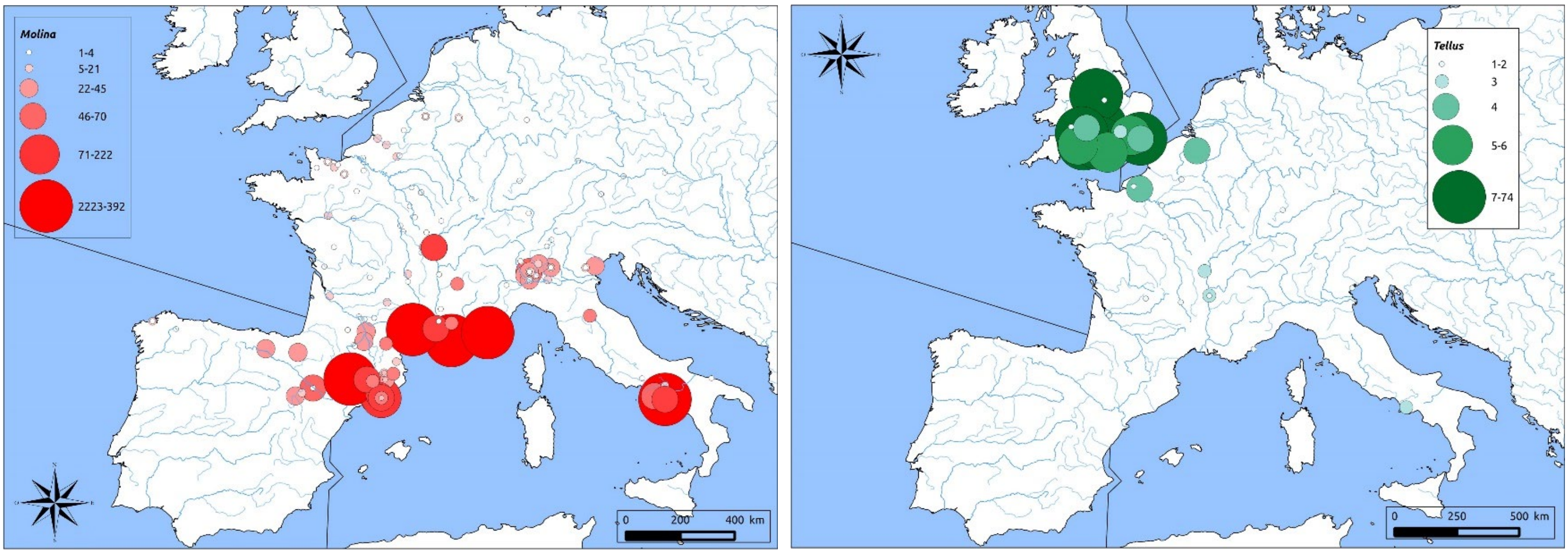

**Fig. 16a-b:** Geographical distribution of *molina* and *tellus* within the CEMA.

[25] Florian Mazel, *Féodalités, 888-1180 […]*, op.cit.

[26] Nicolas Perreaux, « L'écriture du monde (I) […] », *op.cit.*

However, the digitisation of medieval texts makes it possible to go further than these specific impressions, by offering the possibility of comparing the overall evolution of the lexicon of one or several scriptoria. On this graph, for example, each point corresponds to a charter of the famous Abbey of Cluny, characterised by the entire lexicon it contains. A factorial analysis carried out on the whole of this lexicon thus makes it possible to observe the lexical dynamics of the scriptorium of the abbey. We can thus identify the periods in which the writing of acts evolved most strongly, the breaks, the concentration or, conversely, the variability of the forms. Here, we see that the Cluniac vocabulary is initially relatively homogeneous until the first quarter of the 11$^{th}$ century. Then the acts evolve and, above all, diversify strongly in the 11$^{th}$ and especially the 12$^{th}$ century. Finally, they evolve again rapidly, but also become more coherent in the 13$^{th}$ century, on new lexical bases.

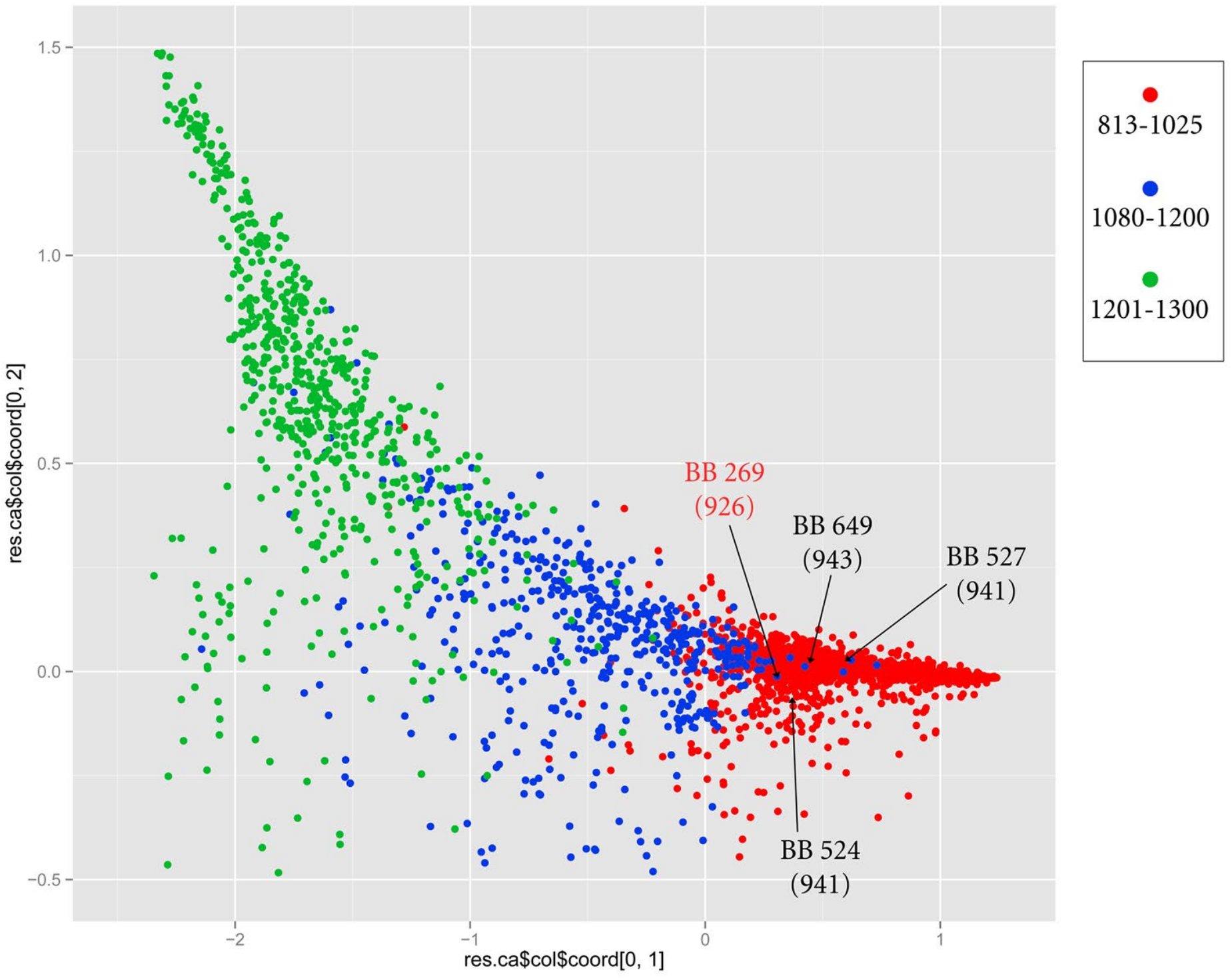

**Fig. 17:** Charters of Cluny: global evolution of the lexicon, 9$^{th}$-13$^{th}$ centuries.

These global methods can of course help to measure regional variability, for example by comparing the lexicon of different settlements. In this experiment, texts from settlements in southern Burgundy were compared with those from settlements in the Loire[27]. For the 11$^{th}$ century, it was thus possible to note that the acts of Cluny and those of Marmoutier (above), often compared in historiography for reasons of individual circulation and institutional links, are in fact quite different. The lexical overlap is indeed very small. Conversely, if we compare Marmoutier to other ecclesiastical institutions in the Loire (Saint-Aubin in Angers, Saint-Julien in Tours, Saint-Florent in Saumur), we do not observe a clear lexical separation. This means that these establishments have a common formulary background. In the same way, the separation between Cluny and Saint-Vincent de Mâcon, although existing, remains rather weak - in particular if one brings it back to the opposition between Marmoutier and Cluny. More specifically, in another article I am writing with Sébastien Barret, we have observed various phenomena of regionalisation in the acts of the long tenth century[28]. One could of course generalise these experiments, and thus map not only lexical, but also social, trends in different areas of medieval Europe - something I would like to devote some time to in the coming years.

[27] Id., « Le rythme de l'écriture. Productions des chartes et dynamique sociale (IX$^{e}$-XIII$^{e}$ siècles) : Bourgogne, Centre, Pays de la Loire », in Chantal Senséby (dir.), *L'écrit monastique dans l'espace ligérien*, Presses Universitaires de Rennes, Rennes, 2018, p. 29-50.

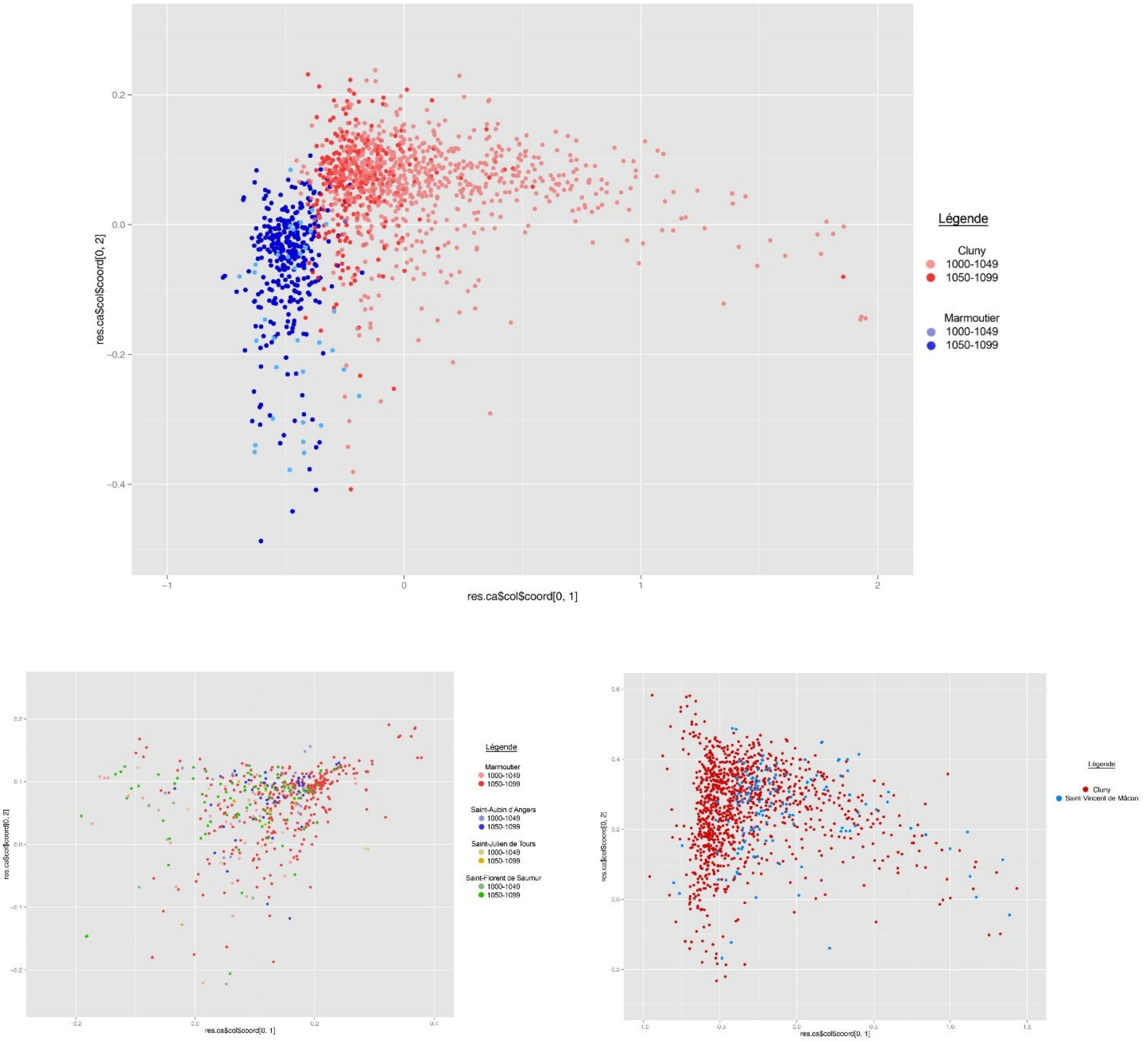

**Fig. 18a-c:** Global comparison of the lexicon, in different corpora for the 11th century. (a) Cluny vs. Marmoutier; (b) Martmoutier vs. Saint-Aubin d'Angers vs. Saint-Julien de Tours vs. Saint-Florent de Saumur; (c) Cluny vs. Saint-Vincent de Mâcon. While there are few differences between the regional corpora (South Burgundy on the one hand: Cluny and Saint-Vincent de Mâcon; "Pays ligériens" on the other), there are important differences between the two areas (Cluny versus Marmoutier).

[28] Sébastien Barret, Nicolas Perreaux, « Le corpus des actes privés en Bourgogne (880-1030). Structure, caractères et sens d'une régionalisation inégale » (forthcoming).

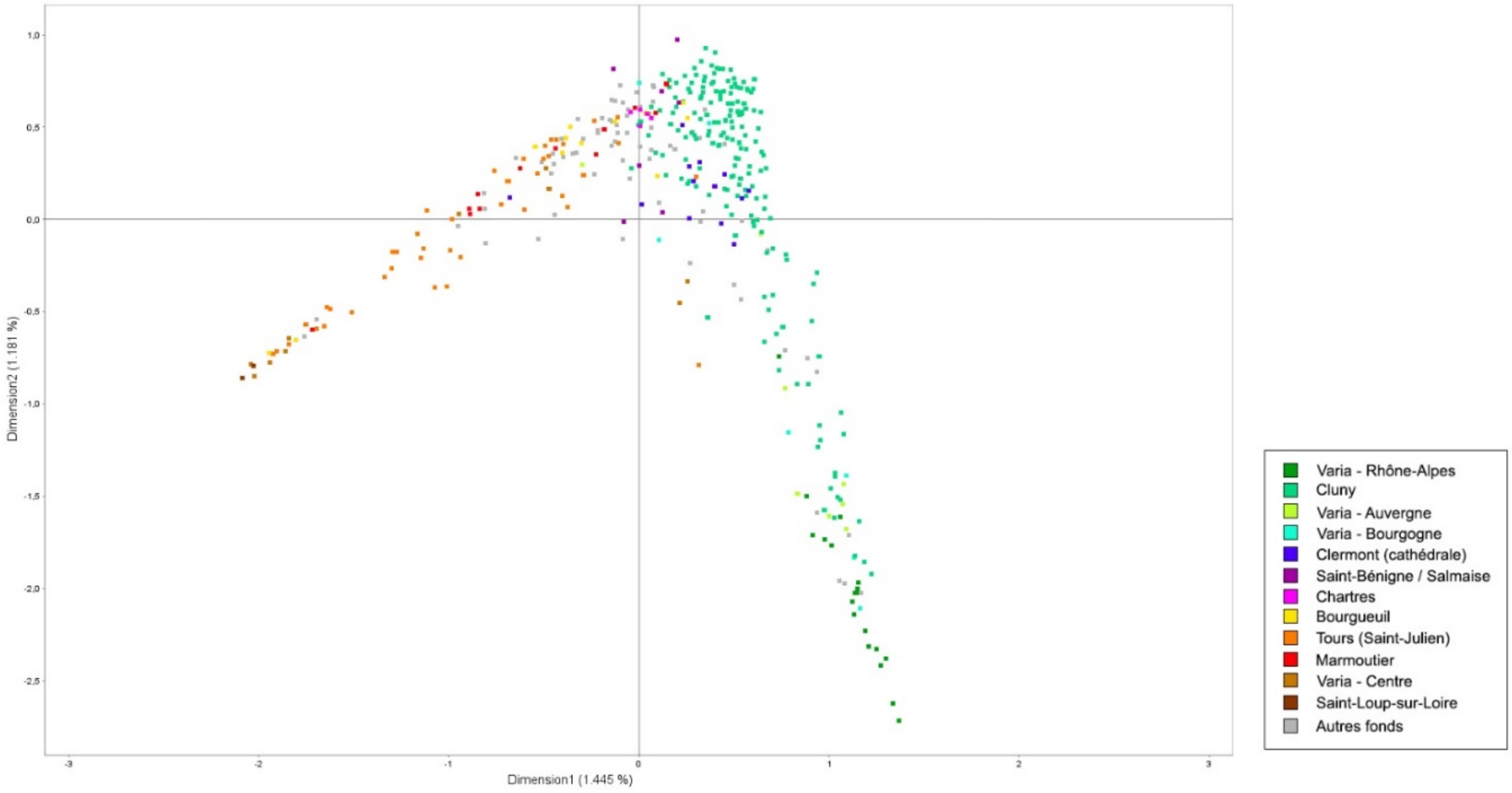

**Fig. 19:** Global analysis of the lexicon in the original acts from different areas/settlements during the long 10th century (factorial analysis of 3220 bi-lemas and 392 acts; each point corresponds to one act).

Of course, the interest of these methods of exploration and modelling of lexical evolutions is not limited to the problem of regionalisation and also directly affects the question of chronological evolutions, on the side of social transformations, but also of written practices. An analysis of the lexicon of Merovingian and Carolingian diplomas, for example, reveals phenomena invisible to the naked eye[29]: 1) the lexicon of diplomas evolves very strongly over the entire period under consideration, particularly if we compare it to the capitularies, which themselves have a very homogeneous lexicon; 2) there is an extremely important break after 800, in particular after the reign of Charlemagne - which thus constitutes a diplomatic in-between. Such experiments could be generalised: they would make it possible to ask different questions, either new or from a new angle. This is once again the case for the question of pontifical documents, which I only briefly touch on here, but which makes it possible to confirm certain historiographical hypotheses on the evolution of writing in this chancery, as well as to propose unpublished or even contradictory observations.

[29] Nicolas Perreaux, « Langue des capitulaires et langue des chartes […] », *op.cit.*

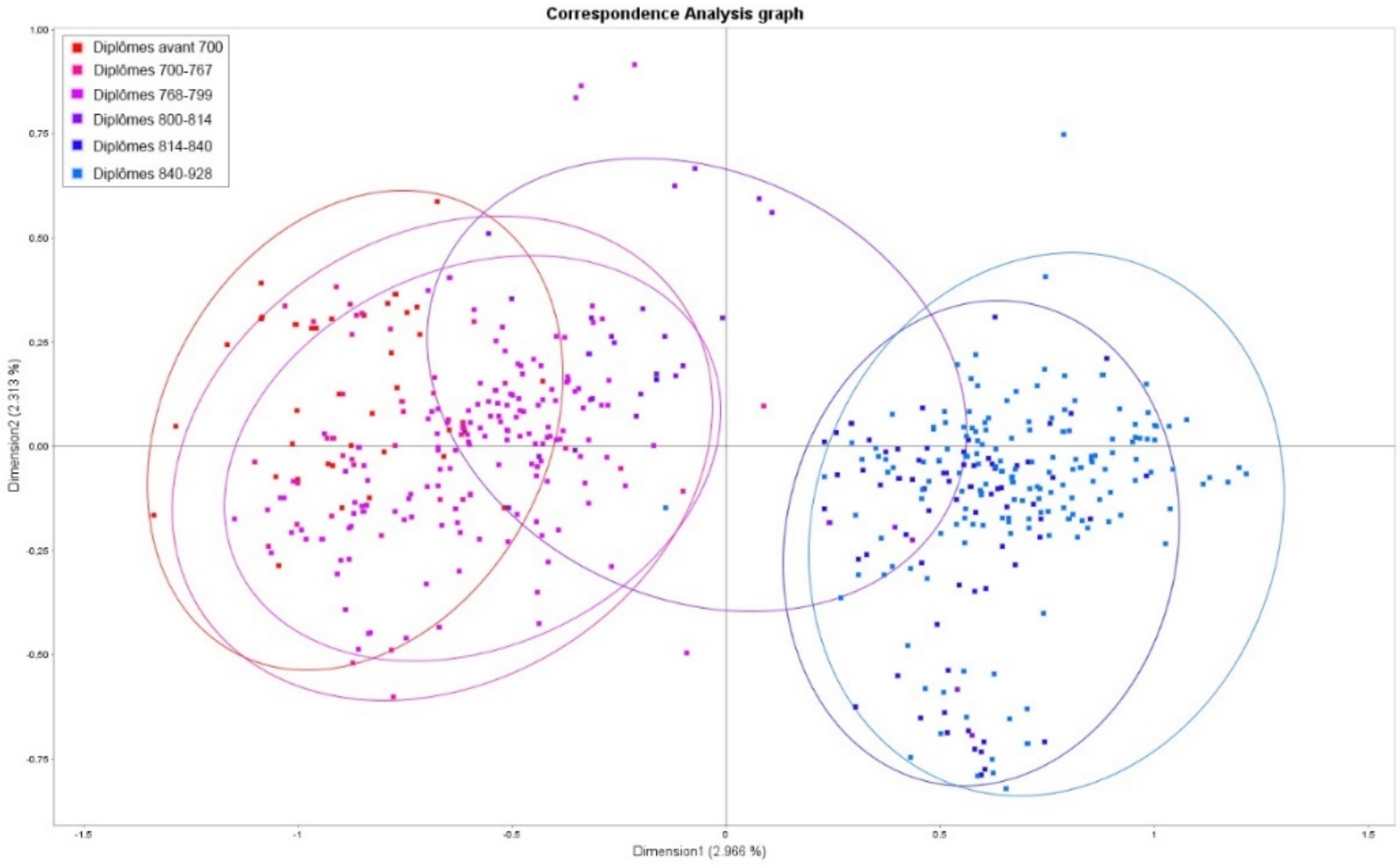

**Fig. 20:** Lexicon of the Merovingian, Carolingian and Post-Carolingian diplomas (dMGH).

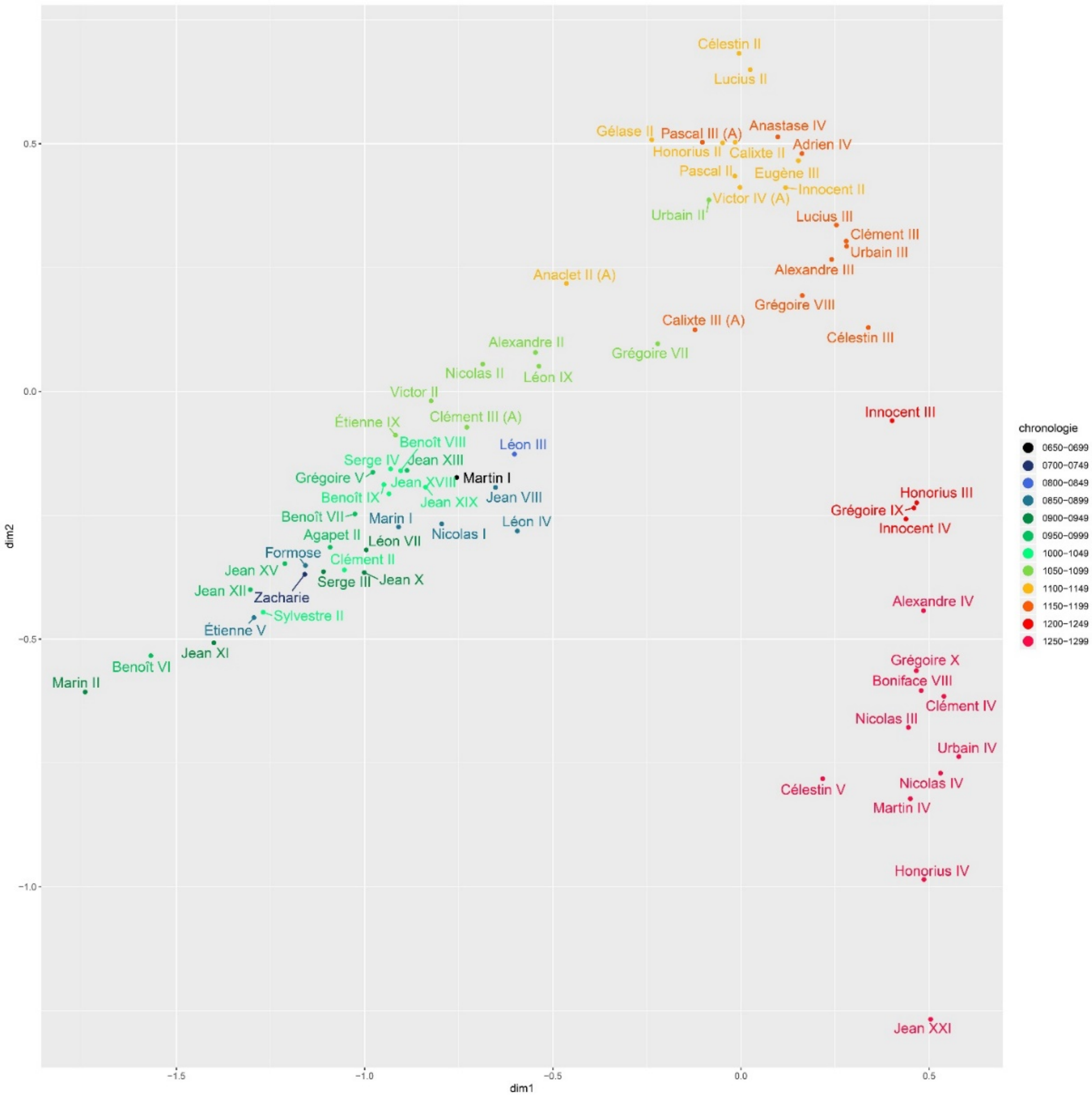

**Fig. 21:** Evolution of the lexicon of pontifical documents in the CEMA, 7th-13th century, by pontificate.

## III. Historical semantics: texts, concepts, things

One last field in which the CEMA could prove valuable is that of historical semantics. The corpora finally make it possible to properly explore this question of the meaning of ancient terms, in this case medieval Latin, in diachronic terms. The potential is undoubtedly considerable. I will mention only a few stages in this historiography. After the founding texts of Humbolt and Herder, the contribution of Jost Trier should be mentioned[30]. He proposed a duo of concepts: field of words and field of meaning. In his publications he showed that the logic of socio-historical structures could be accessed by observing the evolution of terms around pivotal words - in other words, the evolution of co-occurrents. However, his contribution was generally poorly received, for historical as well as technical reasons. Methodologically, Trier did not really explain how to proceed - which was in fact almost impossible without a computer. In 2008, the work of the ANR Omnia made it possible to establish the first medieval lemmatizer, while at the same time proposing a series of tools for the processing of medieval texts[31]. It is in this context that the CEMA appear to be of another possible use. Two files have been selected here to illustrate the approach.

### Se souvenir des belles choses: *Writing, remembering, preserving*

The first is *memoria*[32]. This is a term that is very frequently used in historiography, so much so that Joachim Wollasch wrote as early as 1979 that all medieval sources were commemorative[33]. The German school, in particular the Münster team, played an important role in the dissemination of the concept, which was then passed on to British and American colleagues, then returned to France. Paradoxically, despite this extensive use, semantic studies of the term *memoria* have remained almost non-existent.

Looking at the frequency of the term's evolution, we note first of all that the presence of *memoria* keeps increasing throughout the corpus of the *Patrologia Latina* (PL). A plausible hypothesis is therefore that the theme gains in importance in the theological writings, which make up the bulk of this corpus, between the 3rd and the beginning of the 13th century. In the CEMA charters, the situation is more variable: first, there is a relatively high number of mentions before the ninth century, followed by a drop in the 10th-11th centuries, and then a radical increase after 1050. A possible explanation for the dip in the 10th-11th centuries is probably the proportion of diplomas in the early Middle Ages - the lemma being initially more present in this documentation. This corpus bias corrected (soon, thanks to the metadata provided by artificial intelligence), the trend would probably also be a more or less linear increase, as in the PL.

---

[30] Jost Trier, *Der deutsche Wortschatz im Sinnbezirk des Verstandes : von den Anfängen bis zum Beginn des 13. Jahrhunderts*, Heidelberg, C. Winter, 1931; Lothar Schmidt (dir.), *Wortfeldforschung. Zur geschichte und theorie des sprachlichen feldes*, Darmstadt, Wissenschafliche Buchgesellschaft, 1973; Jost Trier, Anthony Van der Lee, Oskar Reichmann (dir.), *Aufsätze und Vorträge zur Wortfeldtheorie*, Den Haag, Mouton, 1973.

[31] Led by Alain Guerreau. Some of the team's results can be seen at the following address: https://glossaria.eu. The project is currently being developed as a new ANR, Velum, led by Bruno Bon (Paris, IRHT).

[32] Nicolas Perreaux, « An Imaginary Frailty? Memory Discourses in Charters (7th-13th c.) », in Lucie Doležalová, Jan Čermák (dir.), *Discourses and practices of Memory in the Middle Ages*, Brepols, Turnhout, 2021 (forthcoming).

[33] Joachim Wollasch, « Les obituaires, témoins de la vie clunisienne », *Cahiers de civilisation médiévale*, 22-86, 1979, p. 139-171.

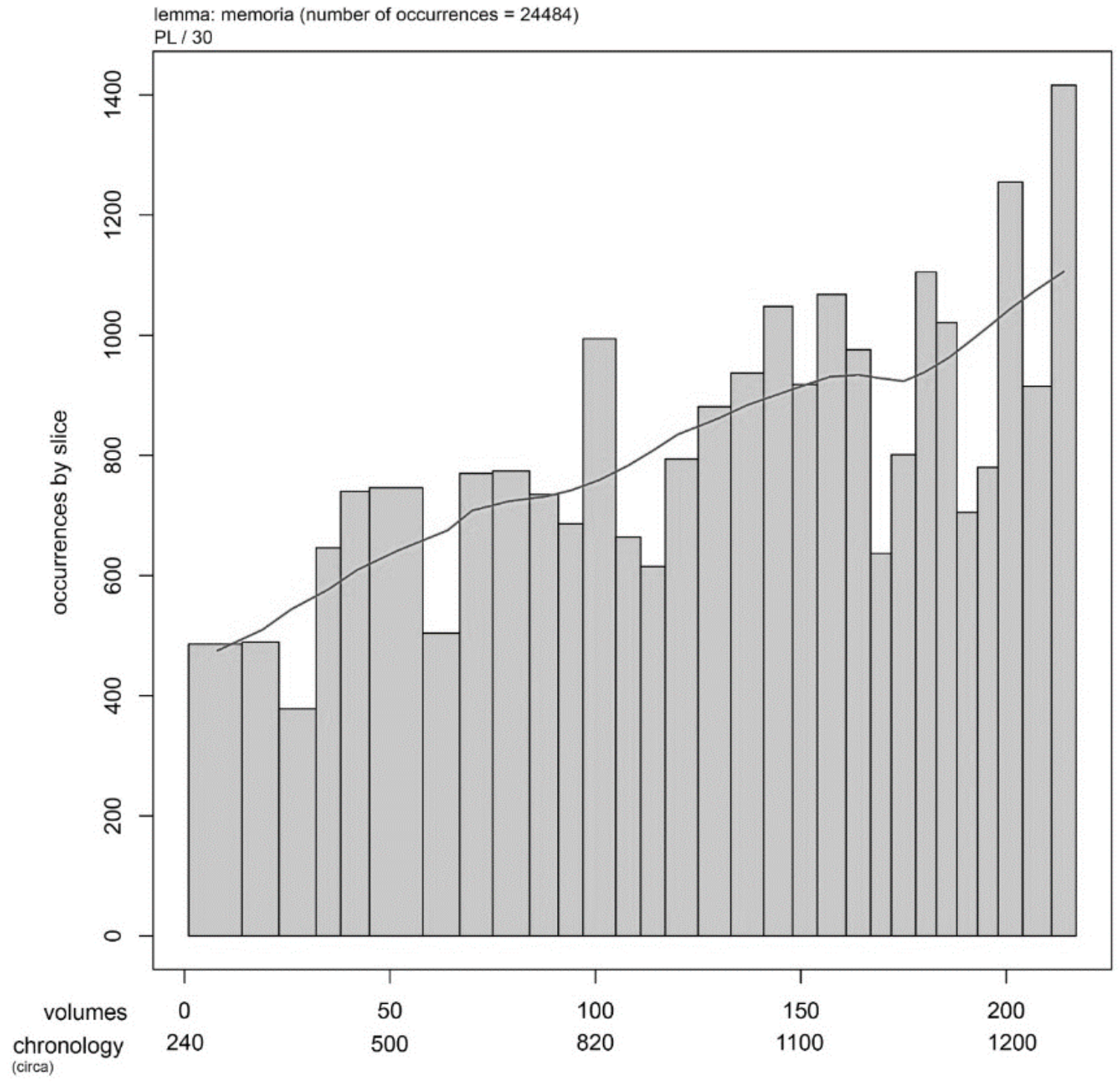

**Fig. 22:** Chronology of the 24,343 mentions of the lemma *memoria* in the PL (3rd-13th century).

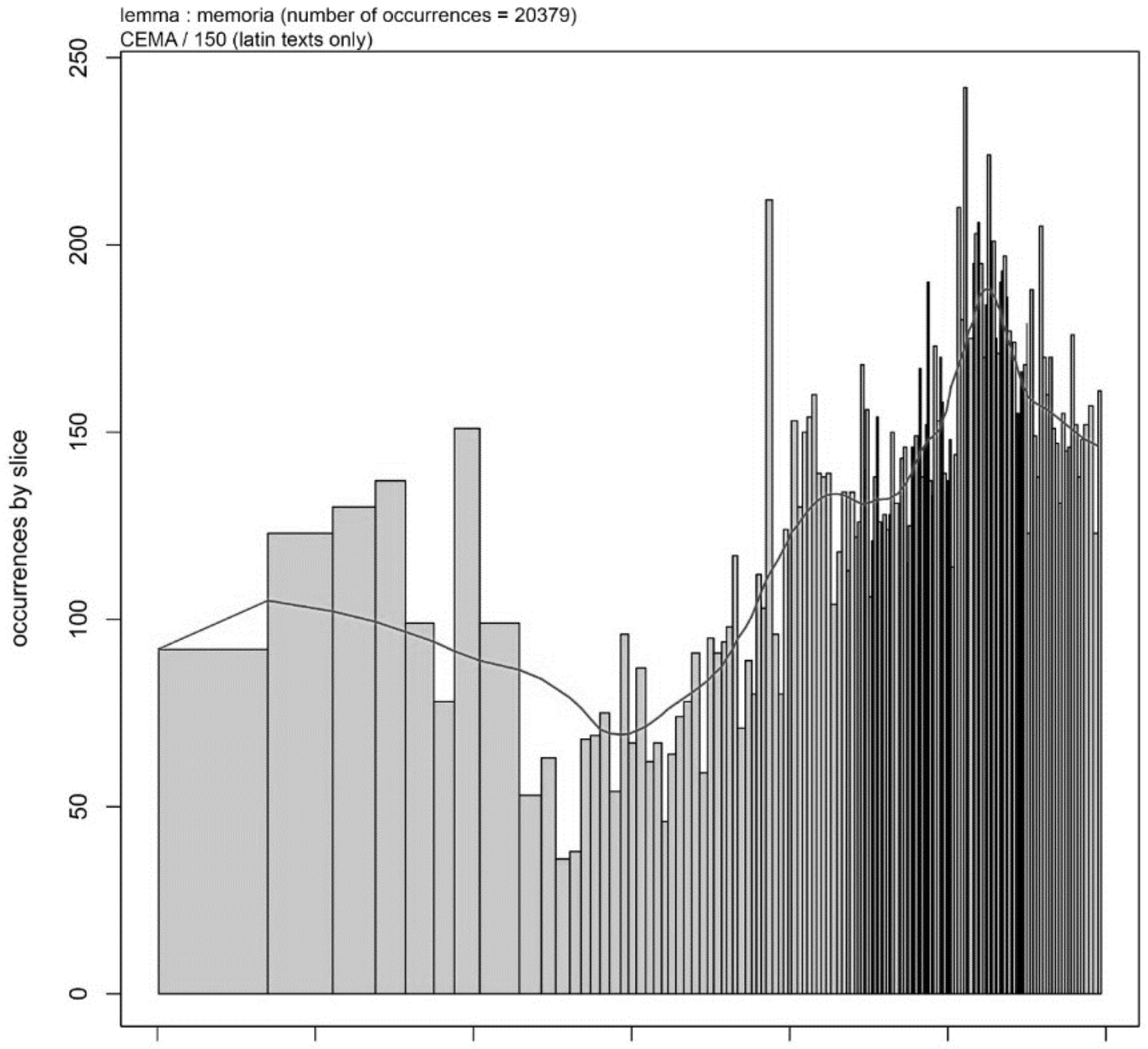

**Fig. 23:** Chronology of the 14,792 mentions of the lemma *memoria* in the CEMA (700-1300).

Let us now turn to the study of the main co-occurrents. In the charters, the terms that appear around *memoria* can be divided into five semantic categories: 1. The good; 2. Oblivion; 3. Time; 4. Ancestors; 5. The written word. We can see that *memoria* can appear in both positive (*bonus*, *felix*, *clarus*, *beatus*, *divus*) and negative contexts (*labilis*, *oblivio*, *fugax*, *evanesco*). But also that the positive or negative characteristics of memory seem to be linked to the medium used for remembrance: the written word or ancestors. In this framework, temporalities represent a kind of background, again between two very marked poles: the linearity of the *mundus* and its decay, or the stability of eternity.

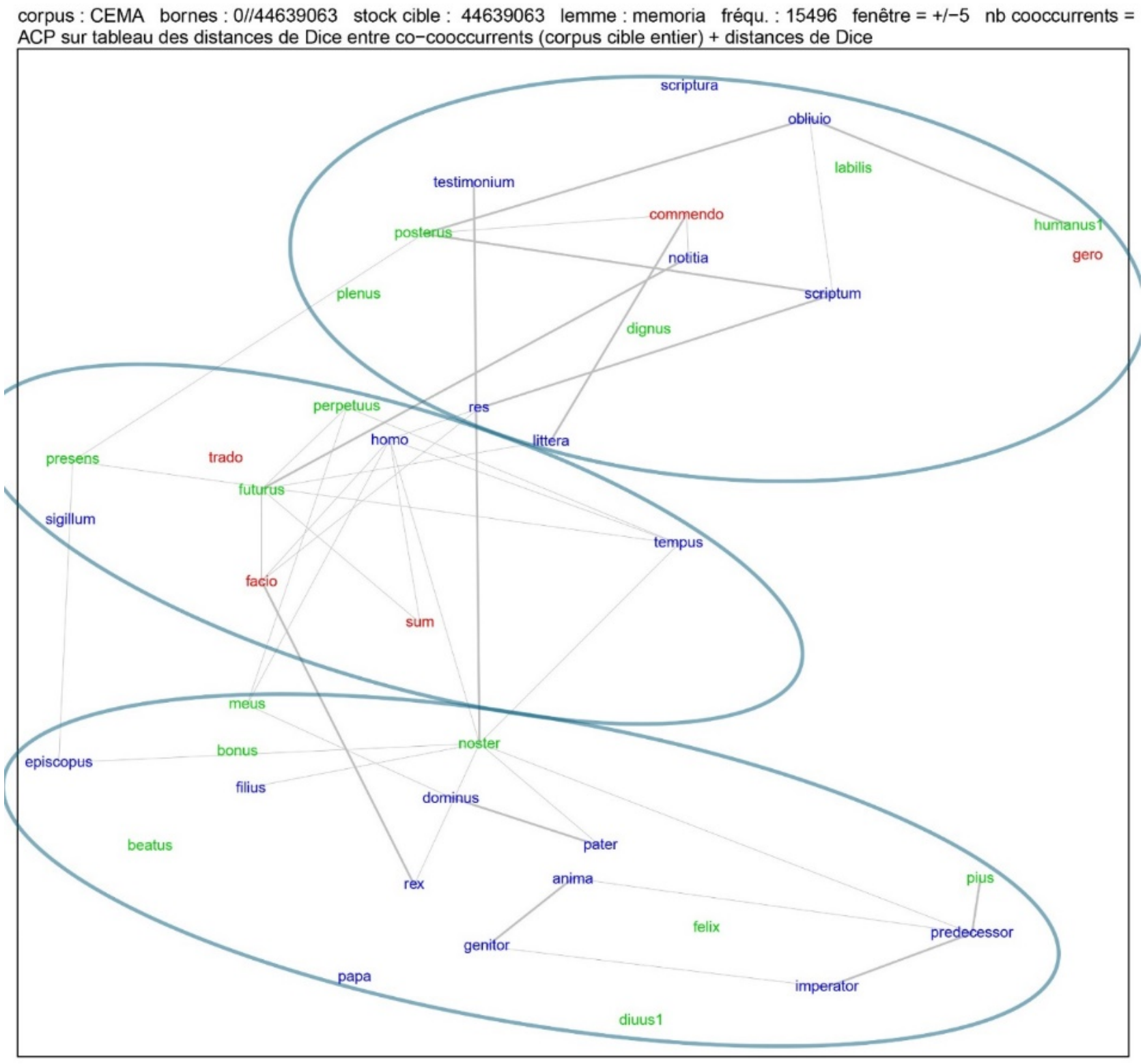

**Fig. 24:** Visualisation of the semantic field around *memoria*, in the CEMA (7th-14th c.). Method: CWB, R, R:CoocB.

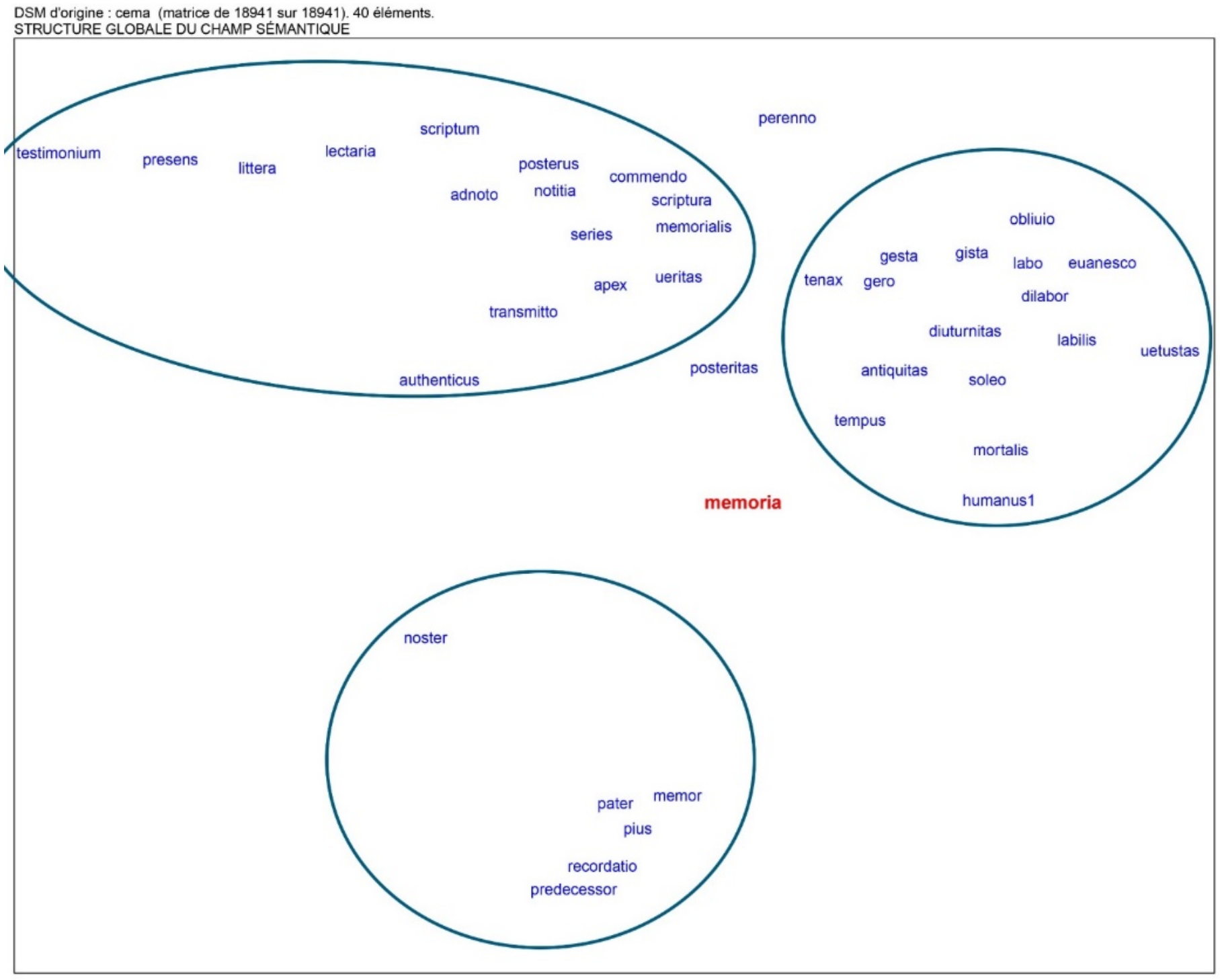

**Fig. 25:** Visualisation of the semantic field around *memoria*, in the CEMA (7th-14th c.). Method: CWB, R, Wordspace, Cooc:WSDSM.

There are several ways of modelling the structure of *memoria* co-occurrences. One way is to see how these terms are articulated by their own cooccurrents (this is called second-order cooccurrence, or 'cooccurrents of cooccurrents'). It shows that these are organised between three poles: 1) At the bottom, who or what is remembered; 2) In the centre, the question of time; 3) Finally, at the top, how remembrance is achieved, with both the problems of the written word and of human fragility. A second method, this time linked to the Wordspace package[34], detects terms appearing in sequences where there are words similar to those surrounding *memoria*. In other words, synonymous contexts, or if you prefer, semantic vectors. We distinguish three sets: writing, time and oblivion, and predecessors. I note in passing that these different methods are complementary: they reveal common elements, but also different tendencies, which need to be re-articulated.

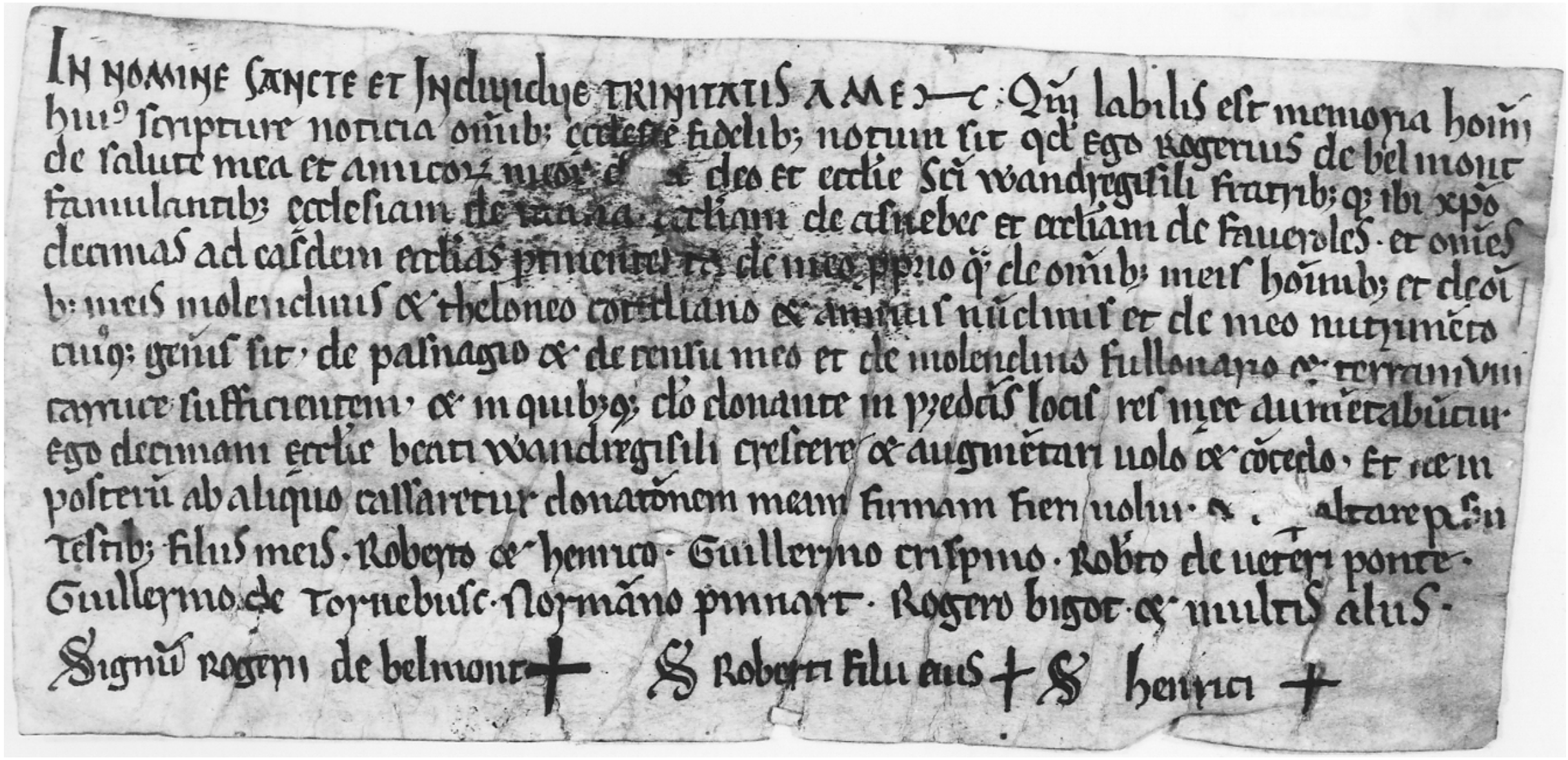

**Fig. 26:** Roger de Beaumont makes a donation to the abbey of Saint-Wandrille (1087, Artem 2721 – picture by the Artem, Nancy).

A qualitative look at the documents thus opens up interesting perspectives. I will take just one example: this original document from the abbey of Saint-Wandrille, which summarises some of the elements noted so far. In this document, we note that the meaning of *memoria* is strongly oriented by the written medium: the memory of men is weak and labile, and that is why another *memoria*, solid and validated, is used via the written word. Thus, the problem is not only what is remembered, but above all how. Medieval discourses insist on the weakness of embodied memory, while spiritual memory - of which clerics are the guarantors – is valued.

A study of the evolution of the cooccurrents – which I will cover very quickly – shows that this situation is reinforced as the chronology progresses. The association between *memoria* and tempus is indeed clearly reinforced in the charters, accentuating the interplay between stability/fragility. The overall evolution of *memoria* is thus towards a form of spiritualisation. All the elements relating to ancestors disappeared over time, while the emphasis on the church and its representatives, written memory, community, forgetting vs. remembering, was constantly reinforced. The memory that 'wins', the good memory, is not that of individuals or ancestors, which is a memory that dies and disappears in time, but the spiritual memory of monks and clerics, which is eternal.

[34] The Wordspace package is developed by Stefan Evert: see Id., « Distributional semantics in R with the wordspace package », *Proceedings of COLING 2014, the 25th International Conference on Computational Linguistics: System Demonstrations*, 2014, p. 110-114.

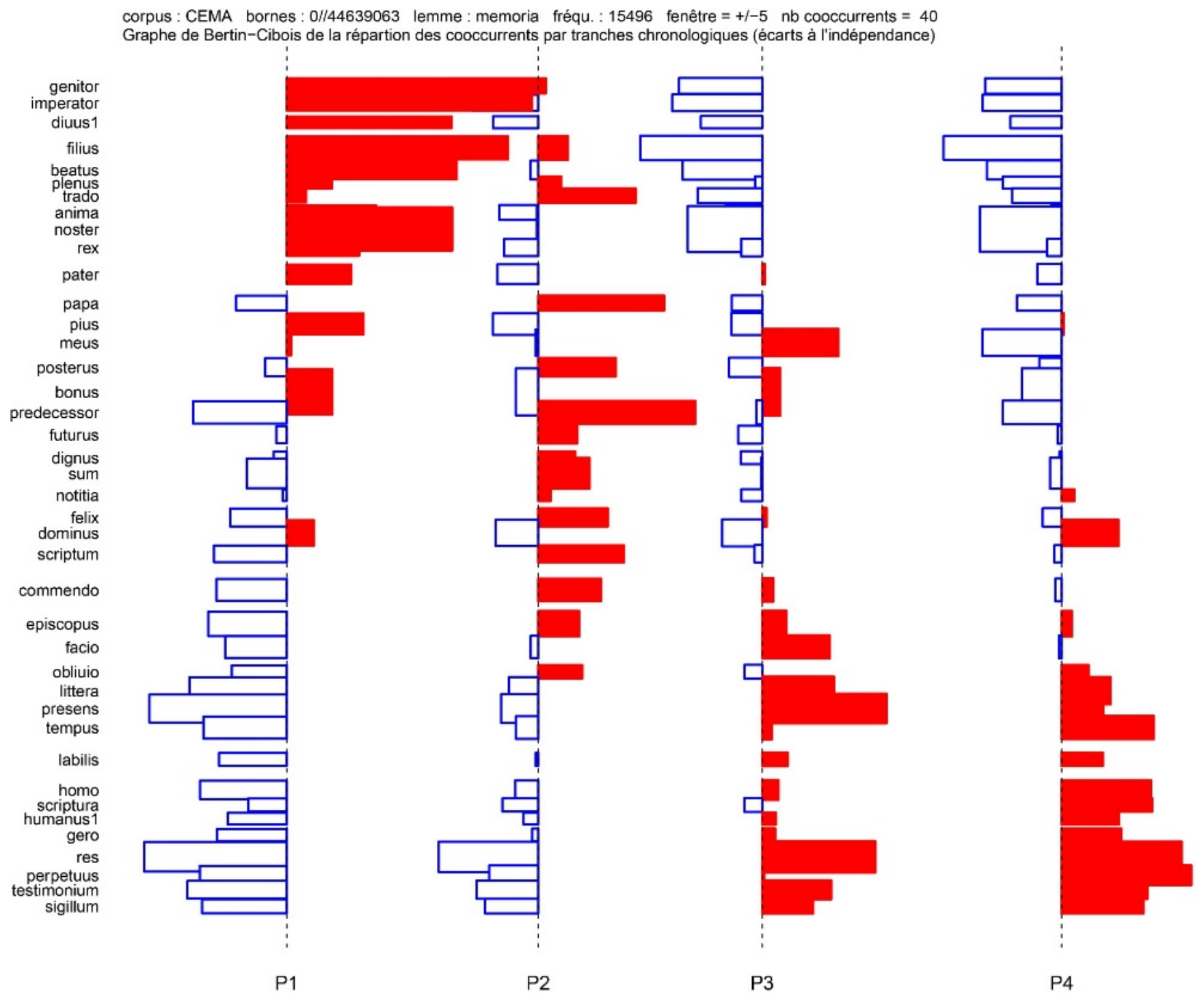

**Fig. 27:** Overall evolution of *memoria* cooccurrents in CEMA, according to four main periods (P1-P4).

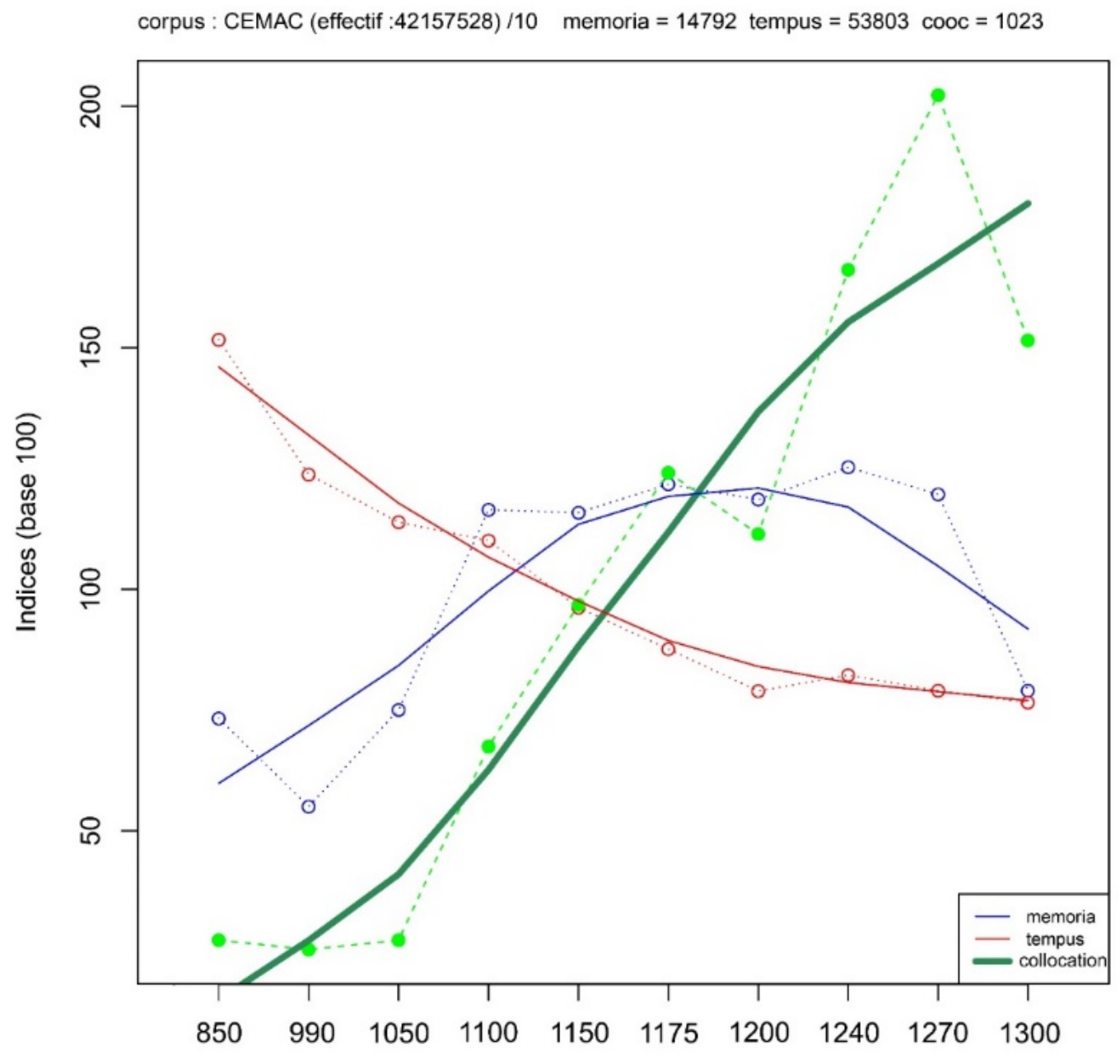

**Fig. 28:** Evolution of the associations between *memoria* and *tempus* over the period 850-1300, in the CEMA.

## *A reformation lexicon?*

the interest of the CEMA therefore also lies in the comparison it permits with corpora of comparable size, including the PL. The second survey presented will concern the so-called ‘reformer’ lexicon[35]. Following Julia Barrow and Laurent Morelle, I proposed a study on the semantics of this lexicon – *corrigo*, *correctio*, *reparo*, *reparatio*, *reformo*, *reformatio*, *emendo*, *emendatio*, etc. – in the context of a study of the ‘reform’ lexicon. This reflection is part of renewed interest in the concept of ‘reform’ and ‘Gregorian reform’ among medievalists. When do we find this lexicon? What does it mean? Does it actually refer to a putative ‘reform’?

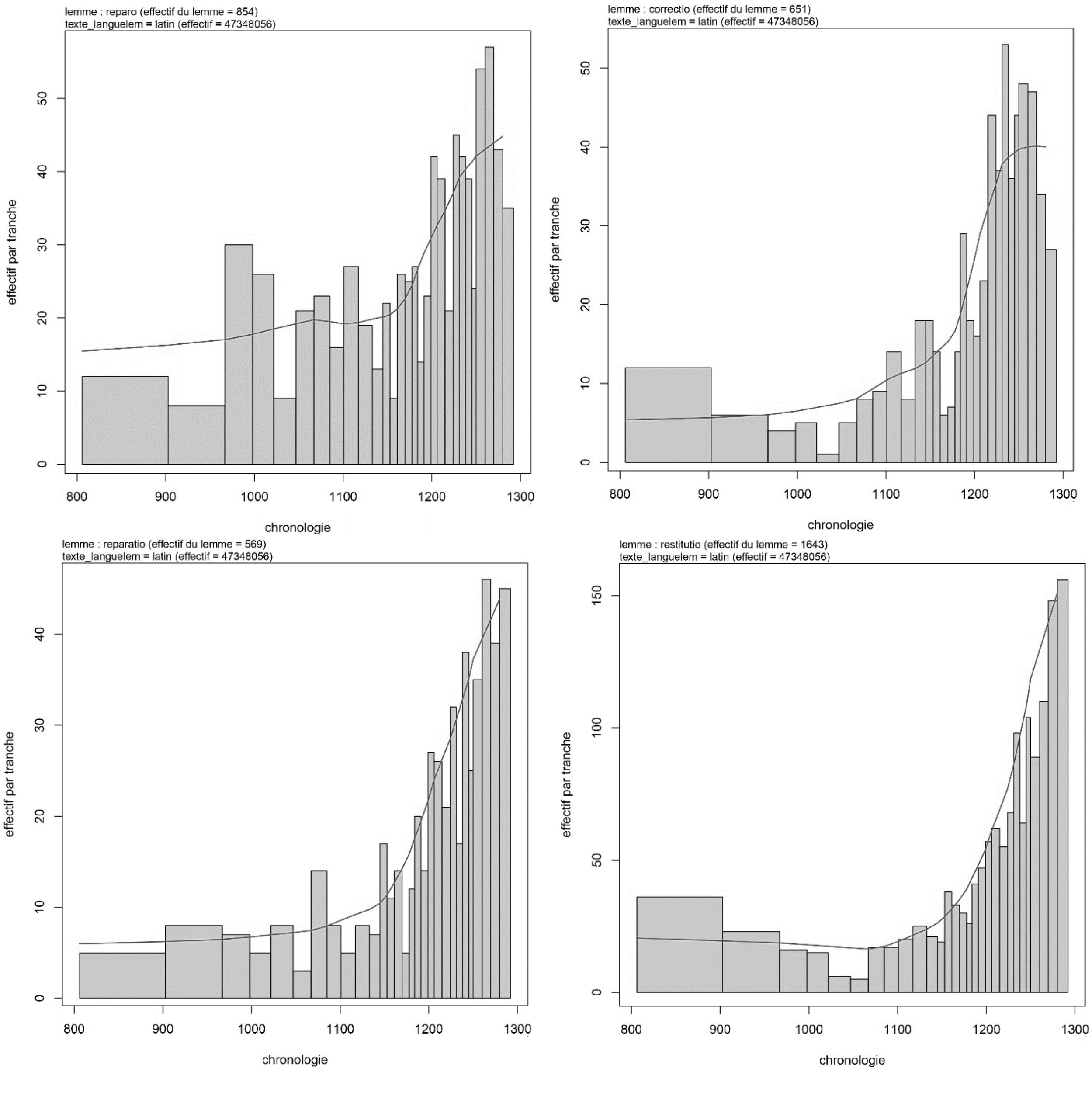

**Fig. 29a-d:** Chronology of the lemmas *reparo*, *correctio*, *restitution*, *reparatio* in the CEMA (800-1300).

[35] Nicolas Perreaux, « Après la Chute, reformer le monde. Réflexions sur la sémantique du lexique dit “réformateur” », in Marie Dejoux (dir.), *Reformatio ? Dire la réforme au Moyen Âge*, Paris, Presses universitaires de la Sorbonne (forthcoming).

Several elements emerged from these surveys: firstly, from a list of 16 so-called 'reformist' terms, it was possible to see that this was a moderately frequent lexicon (32,000 occurrences in total). Secondly, the disparities between the terms were very large: *corrigo* was 100 times more frequent than *reformatio* or *reformator*. Thirdly, nouns (*reformatio*, *correctio*, *emendatio*) arrive later overall than verbs (*reformo*, *corrigo*, *emendo*) and are less frequent - indicating a gradual conceptualisation process. This is particularly the case for a lemma like *reformatio*, of which there are only a few hundred occurrences in the latest version of the CEMA, with a late arrival - centred on the thirteenth century. The corpus of pontifical documents, here on the right, shows an identical situation for the lemma. These elements invited a study of the lexicon, in particular of *reformo* and *reformatio*, going beyond the historiographical concept of 'reform'.

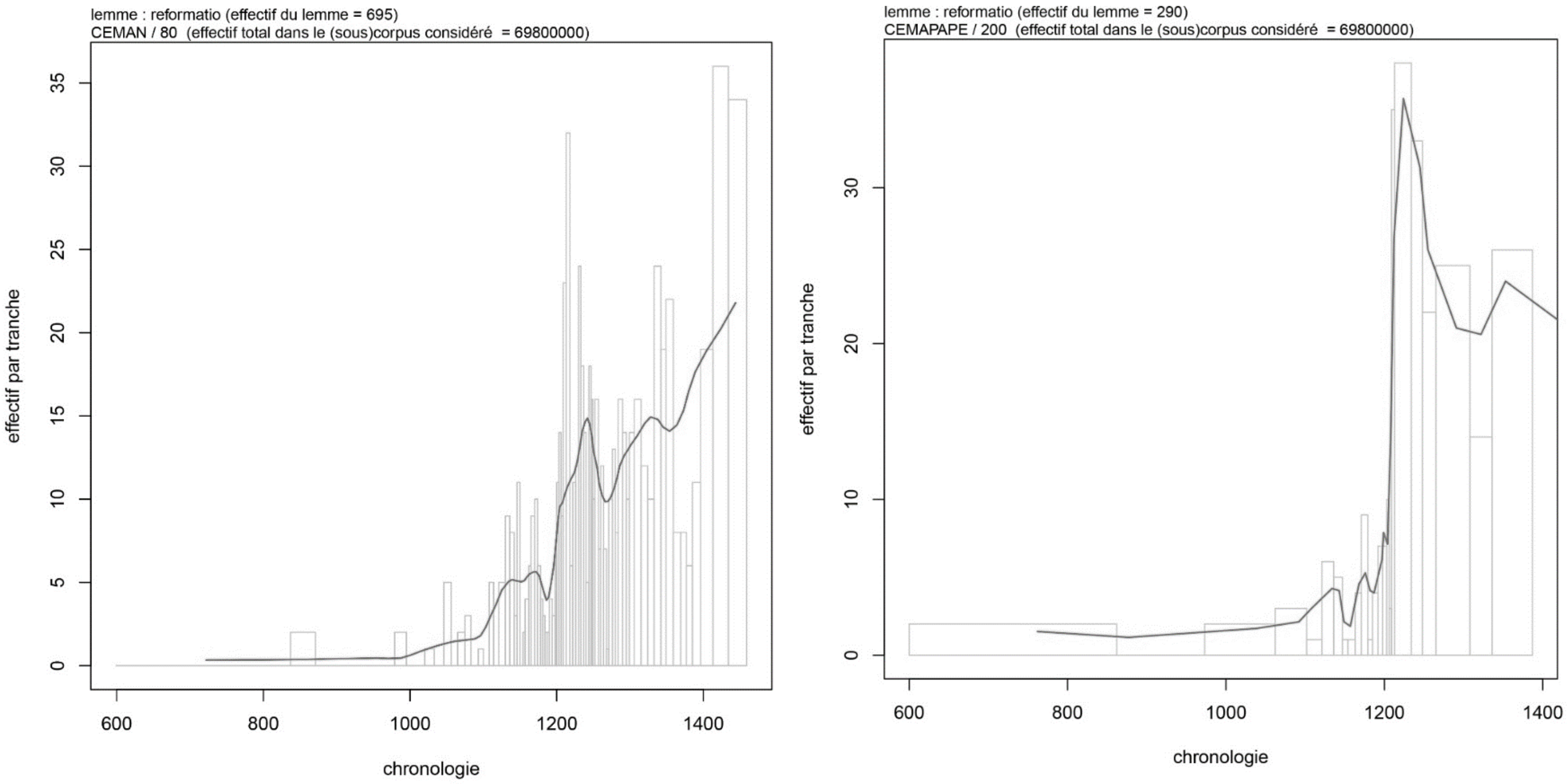

**Fig. 30a-b:** Chronology of the lemmas *reformatio* within the CEMA (left) and in the papal documents of the corpus only (right).

As always, the Vulgate offers key passages for semantic analysis. There are no occurrences of *reformatio*, which is a semi-medieval creation, but two mentions of *reformo*, in the Epistle to the Romans and in the Epistle to the Philippians. Together they invite Christians to reform their minds and bodies, to prepare themselves for the divine plan. The commentary tradition on these verses is very rich, but I will limit myself here to saying that they were extensively taken up and discussed throughout the Middle Ages, and again in the fifteenth century. An investigation into the texts of St Augustine, one of the other main sources for medieval semantics, provides a better understanding of what *reformo* meant. We find elements linked to biblical verses, but also the idea that there are original forms, linked to the Creation (*imago*, *creo*, *prototypus*) - here on the right. The Fall distorted things, degraded them - here on the left - and it is this that we must reform. It seems to me that *reformo* refers to a departure from time, rather than a return to the apostolic ideal.

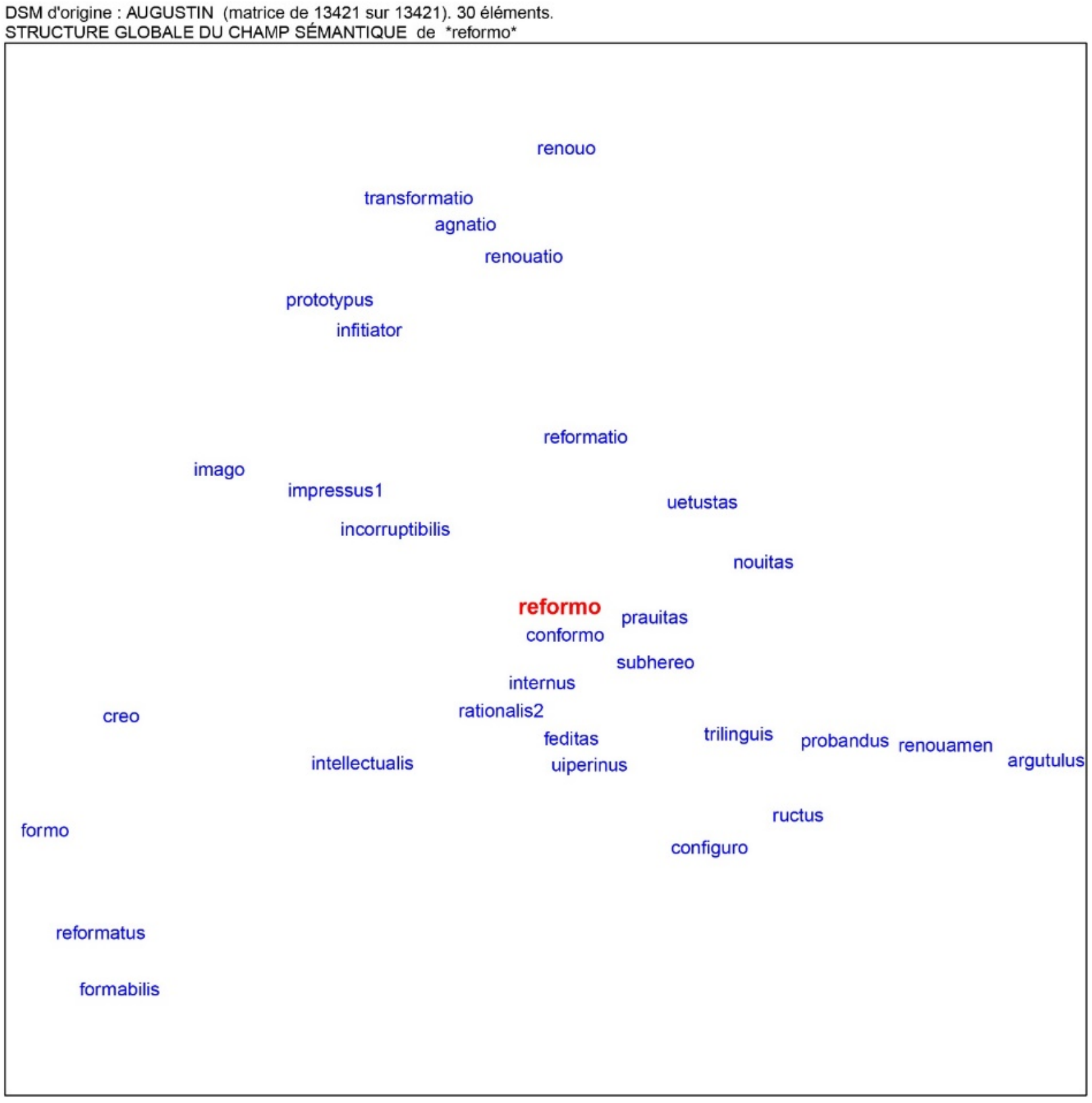

**Fig. 31:** Semantic field of *reformo* in the Augustinian corpus (PL). Method: Wordspace, Rcq, Cooc:WSDSM.

These Augustinian tendencies are visible throughout the PL, even if there are evolutions. In all cases, the articulation is always based on the need to reform something that the century has deformed and distorted from its initial form. I will pass over this very quickly, to concentrate on the evolution of the main cooccurrents of *reformo* and *reformatio* in the PL and the CEMA. The graph presented here is simply an illustration of the proposed synthesis. It seems to me that the logic of *reformo* is gradually extended over the centuries. Initially envisaged in the context of a Christian anthropology by the Fathers, where the aim is to reform the minds of men, the idea of reform gradually extends to bodies, then from bodies to buildings, from buildings to institutions and finally from institutions to the whole of society. These developments are revealed in the evolution of cooccurrents, with terms such as *ecclesia* and *monasterium*, but also later *pax* or *concors*. It is at this time, from the thirteenth century onwards, that the nouns (*reformatio*) develop. The thought of reformation in fact moves from an "action" (verbs) to a thing, a concept applicable to the whole of society. Here again, it is by combining the CEMA with other corpora that certain structures appear.

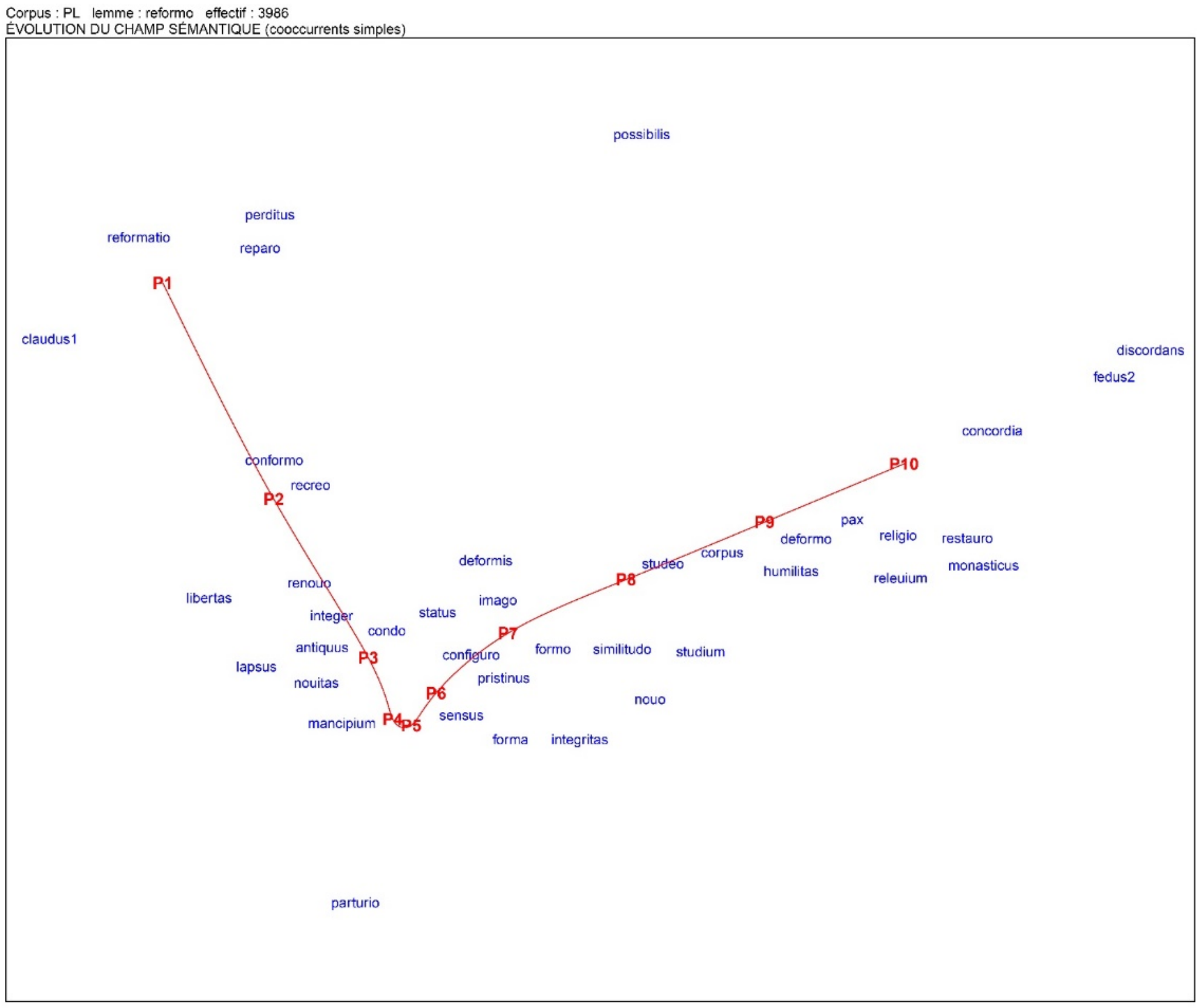

**Fig. 32:** Factorial analysis of the main cooccurrences of *reformo* in the entire PL, according to 10 chronological slices containing an equivalent number of words (P1-P10, i.e. from the Fathers to Innocent III). Method: Rcqp and Cooc:CoocE2.

### *More semantic inquiries: Fatherhood and medieval activities*

More generally, the semantic approach of the CEMA corpus shows to what extent the medieval system was both coherent and different from ours. Approaching the otherness of medieval Europe is indeed complex without computational methods, because qualitative semantics implies, on the contrary, using our own concepts to grasp theirs. Digital technology and corpora free us in part from this limitation (which nevertheless comes back *en force* during the analyses), by allowing us to reconstruct the system of medieval representations 'from the inside'. In order not to multiply the examples, I will present two complementary cases in a synthetic way.

The first concerns the question of medieval fatherhood[36]. "What / who is a father" or rather "What is fatherhood or paternity"? Are they anthropological or even universal structures? We all recall the very famous quote of Darth Vader in the no-less famous 1980 movie *The Empire Strikes Back*: "I am your father". Beyond its iconic dimension, I believe that the scene is perfectly revealing of the current semantics of the term "father". Indeed, for many Western societies at the beginning of the 21st century, the answer to the question "what / who is a father" is relatively unambiguous: according to the *Trésor de la langue française*, the father is first and foremost a "man who has begotten (biologically implied)" or more rarely "who has adopted". This situation seems to have been relatively settled as early as the 18th century. The *Encyclopédie* of Diderot and d'Alembert has different entries for the term: "father (natural right)", "natural father", "legitimate father", "putative father", "adoptive father", "father (sacred criticism)", "conscript father (Roman history)", and finally "fathers of the Church (Hist. ecclésiast.)". The organization of the article leaves little doubt: it is the biological bond of filiation and its implications that takes precedence over everything else, before the juridical claim to fatherhood, the recourse to adoption, and then the meanings linked to the ecclesial system of previous centuries – from the ancient "conscript fathers" or *patres conscripti* to the "sacred criticism" and the "fathers of the Church".

[36] Id., « *In nomine patris*. Éléments pour une sémantique de la paternité médiévale », in Stéphane Lamassé, Octave Julien, Léo Dumont, *Histoire, langues et textométrie*, Paris, Presses universitaires de la Sorbonne, 2021 (forthcoming).

In the case of the medieval system, the answers to these questions are much less unambiguous[37]: the Latin lemma *pater* indeed refers to a multitude of contexts, which still needed to be defined. The investigation in the CEMA and in other digitised corpora, with the help of semantic analysis tools, made it possible to make several observations/conclusions. 1) The semantic field of *pater* and its derivatives underwent an almost total rupture at the turn of Antiquity and the very early Middle Ages. This evolution was intrinsically linked to the establishment of the Christian system, in particular the dogma of the Trinity, which proposed a new definition of the father-son relationship from the Fathers onwards. 2) These observations confirm the strict impossibility of an analysis of medieval kinship that does not recognize a pivotal role for spiritual kinship. The cases of *pater* but also of *paternitas* are edifying in this respect. If one divides the semantic field between a putative "biological" paternity, and secondary symbolic paternities, one irremediably destroys the very meaning of medieval representations and any possibility of reconstructing social structures. 3) The increasing relative importance of spiritual kinship seems to be undeniable. The case of *paternitas* is particularly clear: it is a biblical neologism, at first reserved for God himself, but this quality (a perfect spiritual paternity) gradually extended to popes, bishops, abbots, and ultimately to various key figures in the ecclesial institution. This trend is to be seen in parallel with the multiplication of fathers, a phenomenon already observed by various researchers, and which also extends spiritual fatherhood to all lords / *domini*.

[37] For lexical approaches of *pater*, see Anita Guerreau-Jalabert, « La désignation des relations et des groupes de parenté en latin médiéval », *Archivum Latinitatis Medii Aevi*, vol. XLVI, 1988, p. 65-108; Sylvie Joye, *L'autorité paternelle en Occident à la fin de l'Antiquité et au haut Moyen Âge*, Paris, 2006; Jérôme Baschet, *Le sein du père. Abraham et la paternité dans l'Occident médiéval*, Paris, Gallimard, 2000.

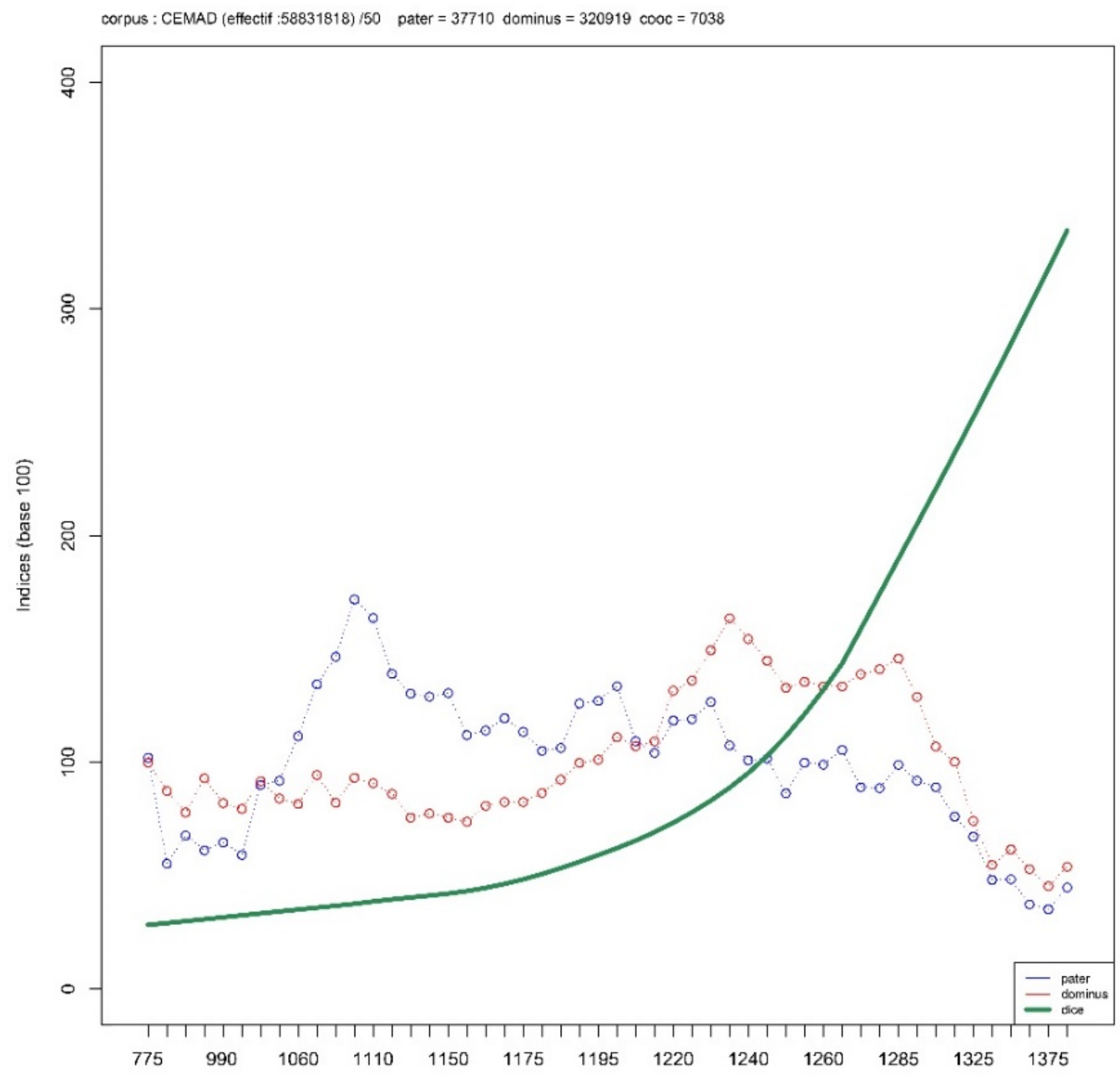

**Fig. 33:** Evolution of the cooccurrences of the lemmas *pater* and *dominus*, in the CEMA (8th-14th centuries). In green: measure of the association, according to the Dice coefficient. Method: Rcqp, Cooc:Freqcooc.

In parallel, the analysis of the semantics of the terms usually seen as relating to 'work' in the Middle Ages (*opus*, *labor*, *servus-servutium*) has shown, on the contrary, that this concept, which is fundamental for contemporary societies (in connection with those of the 'market', the 'wage' and the 'economy'), was in fact strictly incompatible with medieval thought[38]. *Labor* designates a transitory state of suffering, and is in fact carnal (*caro*). Some actions involve labour, but it is not a condition for their proper execution: suffering is not necessary for achievement, it is a consequence of certain activities – a worldly consequence linked to the Fall. Strongly anchored in its century, *labor* is not very polarising. As for *servitium*, it is cyclical, designating duties, actions and iterative functions, which return again and again to manifest the status of some and others, and thus reproduce domination. *Servitium* is not fixed in the strict sense, since it is rhythmic, but it polarises all the same, since it is directed towards lords or God. *Officium* and *beneficium* also seem to fit into this framework. *Opus* designates simultaneously a process and a realisation, in which the latter predominates and remains alone at the end of the chain. The stability and positive valences associated with the term make *opus* an eminently spiritual lemma. The *labeur* is often done for the Church and for the spiritualisation of the world. Basically, *opus* fixes and polarises: one works for something, most often for the salvation of the soul, for the Church. *Ars*, *opifex* and other terms relate to techniques and skills, and refer rather to *opus* – without being perfectly articulated with it.

[38] Nicolas Perreaux, « Œuvrer, servir, souffrir. Réflexions sur la sémantique des activités médiévales », in Michel Lauwers (dir.), *Labeur et production au sein des monastères de l'Occident médiéval*, Turnhout, Brepols, 2021, p. 31-79.

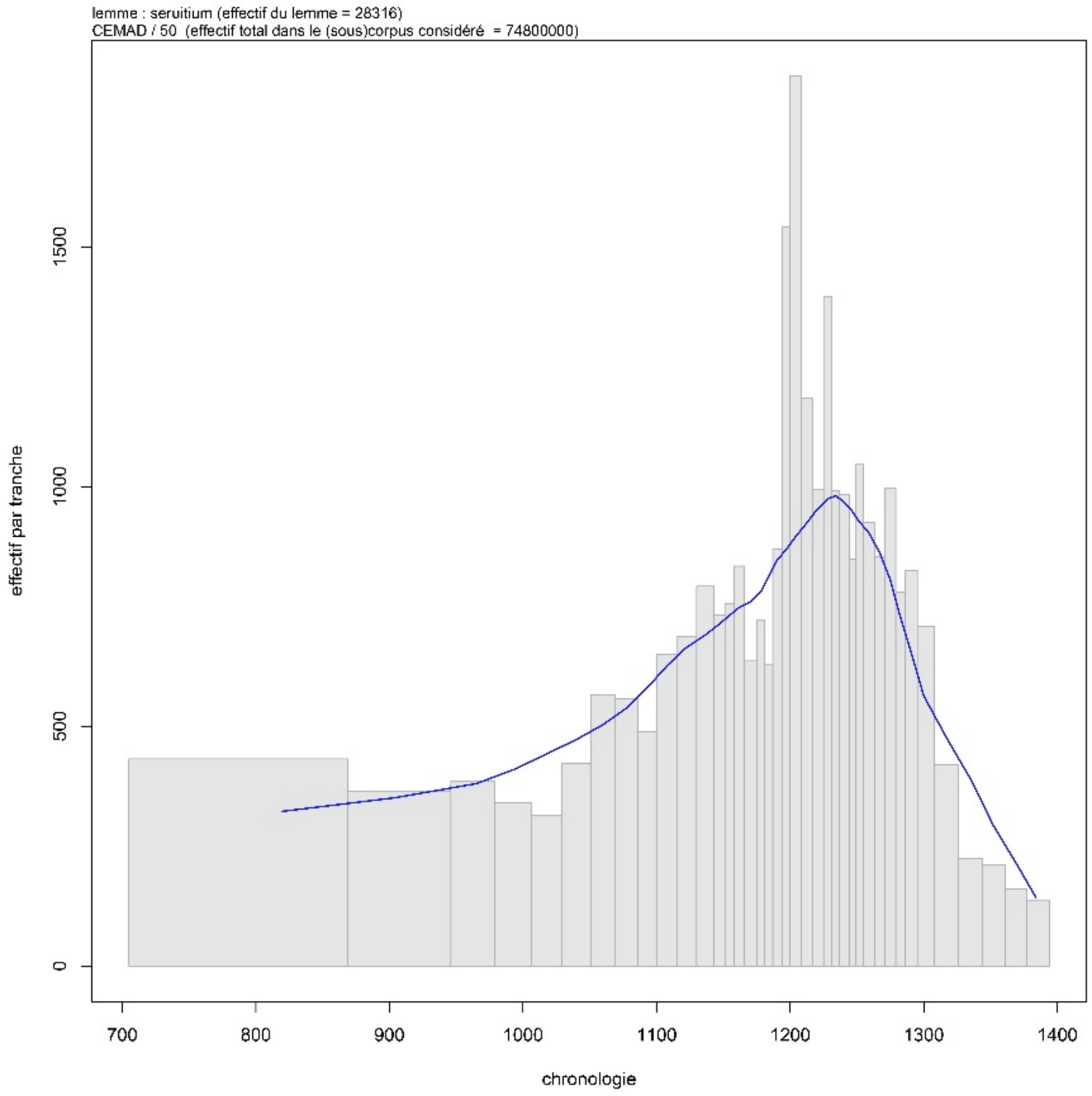

**Fig. 34:** Chronology of *servitium* within the CEMA (7th-14th century).

*

The CEMA corpus could render many services to medievalists: From the analysis of written practices, through the measurement of lexical richness, to semantic studies, via the examination of medieval dynamics and regionalisation. While it certainly does not exclude traditional approaches, it provides a new scale of analysis for many unresolved or unpublished questions. Contrary to what is sometimes argued, I believe that digital technology does not distance us from documentation and materiality. First of all, because it does not 'dematerialise', as we often read. From hard disks to cables, nothing in digital technology is immaterial. The CEMA are no exception to the rule. They are thus part of a very long chain of materiality, made up of productions, conservation gestures and sometimes successive sorting, from the scribes' *calames* to our keyboards and algorithms. Better still, it seems to me that corpora of this size make it possibleto open up investigations into the materiality of documents, provided that they are preserved, indexed and that adequate metadata are produced – with, for example, investigations into the preservation of originals compared to copies, methods of *cartularisation*, the size of documents, etc.

The aim of the corpus, which will be made available in open access in 2021, is therefore to enable and strengthen the analysis of medieval society as a whole, on a European scale[39]. In particular, I would like to emphasise a central interest of the corpus: the articulation of the different scales of analysis. For a long time, medievalists were confined either to a local/regional approach, based on documentation, or to a global approach, based on monographs. For the first time, the corpus makes it possible to carry out complex interrogations, in an ensemble covering Western Europe in an increasingly homogeneous manner, from the very early Middle Ages to the late 15$^{th}$ century – while at the same time favouring investigations into local or regional ensembles. This possibility will make it possible to accumulate data that will probably be contradictory, but in a good way, not only with the current model of medieval history, but between the different scales of analysis. The CEMA will strengthen the understanding of the global specificity of medieval Europe, its overall dynamics, and the nature of its regional variants – the changeability of which was, paradoxically, a formativeelement of the overall coherence and dynamics of the system.

[39] For this perspective, see Marc Bloch, « Pour une histoire comparée des sociétés européennes », *Revue de synthèse historique*, t. 46, 1928, p. 15-50; Georges Duby, *L'économie rurale et la vie des campagnes […]*, op.cit.; Jacques Le Goff, *La civilisation de l'Occident médiéval*, Paris, Arthaud, 1964; Id., *L'Europe est-elle née au Moyen Âge*, Paris, Seuil, 2003; Alain Guerreau, *Le féodalisme, un horizon théorique*, Paris, Le Sycomore, 1980; Robert Fossier, *Enfance de l'Europe […]*, op.cit.; Michael Mitterauer, *Warum Europa? Mittelalterliche Grundlagen eines Sonderwegs*, München, C.H. Beck, 2003; Jérôme Baschet, *La civilisation féodale : de l'an mil à la colonisation de l'Amérique*, Paris, Aubier, 2004; Pierre Toubert, *L'Europe dans sa première croissance. De Charlemagne à l'an mil*, Paris, Fayard, 2004; Chris Wickham, *Framing the Early Middle Ages […]*,

---

op.cit.; Christopher Loveluck, *Northwest Europe in the Early Middle Ages, c. AD 600-1150. A Comparative Archaeology*, Cambridge, Cambridge University Press, 2013.